\documentclass[twocolumn]{aastex62}
\usepackage[section]{placeins}
\usepackage{longtable}
\usepackage{hyperref}
\usepackage{url}
\graphicspath{{./}{figures/}}

\shorttitle{SN\,2019stc}
\shortauthors{Gomez et al.}

\begin{document}

\title{The Luminous and Double-Peaked Type Ic Supernova 2019stc: Evidence for Multiple Energy Sources}

\correspondingauthor{Sebastian Gomez}
\email{sgomez@cfa.harvard.edu}
	
\author[0000-0001-6395-6702]{Sebastian Gomez}
\affil{Center for Astrophysics \textbar{} Harvard \& Smithsonian, 60 Garden Street, Cambridge, MA 02138, USA}

\author[0000-0002-9392-9681]{Edo Berger}
\affil{Center for Astrophysics \textbar{} Harvard \& Smithsonian, 60 Garden Street, Cambridge, MA 02138, USA}

\author[0000-0002-0832-2974]{Griffin Hosseinzadeh}
\affil{Center for Astrophysics \textbar{} Harvard \& Smithsonian, 60 Garden Street, Cambridge, MA 02138, USA}

\author[0000-0003-0526-2248]{Peter K. Blanchard}
\affil{Center for Interdisciplinary Exploration and Research in Astrophysics and Department of Physics and Astronomy, \\Northwestern University, 1800 Sherman Ave, Evanston, IL 60201, USA}

\author[0000-0002-2555-3192]{Matt Nicholl}
\affil{Birmingham Institute for Gravitational Wave Astronomy and School of Physics and Astronomy, University of Birmingham, Birmingham B15 2TT, UK}

\author[0000-0002-5814-4061]{V. Ashley Villar}
\altaffiliation{Simons Junior Fellow}
\affiliation{Department of Astronomy, Columbia University, New York, NY 10027-6601, USA}

\begin{abstract}
We present optical photometry and spectroscopy of SN\,2019stc (=ZTF19acbonaa), an unusual Type Ic supernova (SN Ic) at a redshift of $z=0.117$. SN\,2019stc exhibits a broad double-peaked light curve, with the first peak having an absolute magnitude of $M_r=-20.0$ mag, and the second peak, about 80 rest-frame days later, $M_r=-19.2$ mag. The total radiated energy is large, $E_{\rm rad}\approx 2.5\times 10^{50}$ erg. Despite its large luminosity, approaching those of Type I superluminous supernovae (SLSNe), SN\,2019stc exhibits a typical SN Ic spectrum, bridging the gap between SLSNe and SNe Ic. The spectra indicate the presence of Fe-peak elements, but modeling of the first light curve peak with radioactive heating alone leads to an unusually high nickel mass fraction of $f_{\rm Ni}\approx 0.31$ ($M_{\rm Ni}\approx 3.2$ M$_\odot$). Instead, if we model the first peak with a combined magnetar spin-down and radioactive heating model we find a better match with $M_{\rm ej}\approx 4$ M$_\odot$, a magnetar spin period of $P_{\rm spin}\approx 7.2$ ms and magnetic field of $B\approx 10^{14}$ G, and $f_{\rm Ni}\lesssim 0.2$ (consistent with SNe Ic). The prominent second peak cannot be naturally accommodated with radioactive heating or magnetar spin-down, but instead can be explained as circumstellar interaction with $\approx 0.7$ $M_\odot$ of hydrogen-free material located $\approx 400$ AU from the progenitor. Accounting for the ejecta mass, CSM shell mass, and remnant neutron star mass, we infer a CO core mass prior to explosion of $\approx 6.5$ M$_\odot$. The host galaxy has a metallicity of $\approx 0.26$ Z$_\odot$, low for SNe Ic but consistent with SLSNe. Overall, we find that SN\,2019stc is a transition object between normal SNe Ic and SLSNe.
\end{abstract}

\keywords{supernova: general -- supernova: individual (SN\,2019stc)}

\section{Introduction} 
\label{sec:intro}

Stars more massive than $\approx 8$ M$_\odot$ are expected to end their lives in core collapse supernova (CCSN) explosions. Among these, Type Ic SNe (SNe Ic) are the explosions of stars that have lost their hydrogen and helium envelopes. SNe Ic have typical peak magnitudes of $M_r = -17.7\pm 0.8$ mag (e.g., \citealt{Filippenko97, Lyman16, Prentice16, Fremling18, Prentice19, Barbarino20}), while the subset of broad-lined SNe Ic (SNe Ic-BL) can reach peak magnitudes of up to $\approx -19.4$ mag \citep{Taddia19_broadlined}. It is well established that SNe Ic are powered by the radioactive decay of $^{56}$Ni synthesized in the explosion, with typical nickel masses of $M_{\rm Ni} \approx 0.05-0.6$ M$_\odot$, or nickel fractions of $f_{\rm Ni} \approx 0.05-0.3$ \citep{Drout11,Taddia19_broadlined,Barbarino20}. In terms of their environments, SNe Ic tend to occur in galaxies with relatively high metallicities of \mbox{$12 + \log($O/H$) = 8.8\pm 0.2$} \citep{Modjaz20}.

Type I superluminous supernovae (SLSNe) on the other hand are a much more rare type of CCSNe. SLSNe also represent the explosions of stripped stars, but they be up to 100 times more luminous than SNe Ic, with peak magnitudes of $M_r\approx -20.0$ to $-23$ mag (e.g., \citealt{Chomiuk11,Quimby11,Lunnan13, Gal-Yam19,Gomez20}). Their early time spectra tend to be blue and characterised by oxygen features near $\sim 4000$ \AA\ (e.g, \citealt{Gal-Yam12,Nicholl17_egm}), but their late time spectra are closer to those of SNe Ic at earlier phases (e.g., \citealt{Quimby18, Nicholl19_nebular}). The primary energy source of SLSNe appears to be the spin-down energy of a newly formed millisecond magnetar, with typical spin periods of $P_{\rm spin}\approx 1.2 - 4$ ms and magnetic fields of $B\approx 0.2 - 1.8\times 10 ^{14}$ G \citep{Nicholl17}. More recently, the ejecta masses of SLSNe have been shown to extend to $\approx 40$ M$_\odot$, much higher than for SNe Ic \citep{Blanchard20}. SLSNe are known to be over-represented in low metallicity galaxies, which have typical metallicities of \mbox{$12 + \log($O/H$) = 8.4\pm 0.3$} \citep{Lunnan14}.

CCSNe with luminosities intermediate to SNe Ic and SLSNe, as well as their detailed properties and relation to the other classes, remains mostly unexplored. Only a handful of events with peak magnitudes in the $M_r = -19$ to $-20.0$ range are known (SN\,2018gep \citealt{Ho19}, SN\,2018bsz \citealt{Anderson18}, SN\,2011kl \citealt{Greiner15}, SN\,2012aa \citealt{Roy16}, and DES14C1rhg and DES15C3hav \citealt{Angus19}). Such intermediate events may potentially be powered by radioactivity and/or a magnetar engine, and can therefore shed light on the relation between the energy sources and progenitors of the various types of stripped CCSNe.

Here, we present detailed optical observations of an intriguing event intermediate to SNe Ic and SLSNe. Originally classified as a SLSN-I by \cite{Yan20_2019stc}, SN\,2019stc exhibits a spectrum typical of SNe Ic at all phases of its evolution. Its light curve is peculiar in several ways: it is unusually luminous ($M_r\approx -20$ mag), unusually broad (rise time of 40 rest-frame days), exhibits a prominent second peak ($M_r \approx -19.2$ mag) about 80 rest-frame days after the first peak, and radiated $\approx 2.5\times 10^{50}$ erg. From analytical models we find that the first peak is predominantly powered by a magnetar engine, but with possible contribution from radioactive heating at a level typical of SNe Ic (thus explaining the spectral appearance). The second peak is most likely explained as delayed circumstellar interaction with a shell ejected prior to the SN explosion.

The paper is structured as follows: in \S\ref{sec:observations} we present our classification and follow-up observations of SN\,2019stc, as well as publicly available data. We describe the multi-color and bolometric light curves in \S\ref{sec:analysis}, and the optical spectra in \S\ref{sec:spectra}. In \S\ref{sec:modeling} we present a range of analytical models of the light curve, and in \S\ref{sec:host} we describe the host galaxy properties. In \S\ref{sec:conclusion} we discuss possible interpretations of SN\,2019stc and present our conclusions. Throughout the paper we assume a flat $\Lambda$CDM cosmology with \mbox{$H_{0} = 69.3$ km s$^{-1}$ Mpc$^{-1}$}, $\Omega_{m} = 0.286$, and $\Omega_{\Lambda} = 0.712$ \citep{hinshaw13}.

\section{Observations}
\label{sec:observations}

\begin{figure}[t!]
	\begin{center}
		\includegraphics[width=\columnwidth]{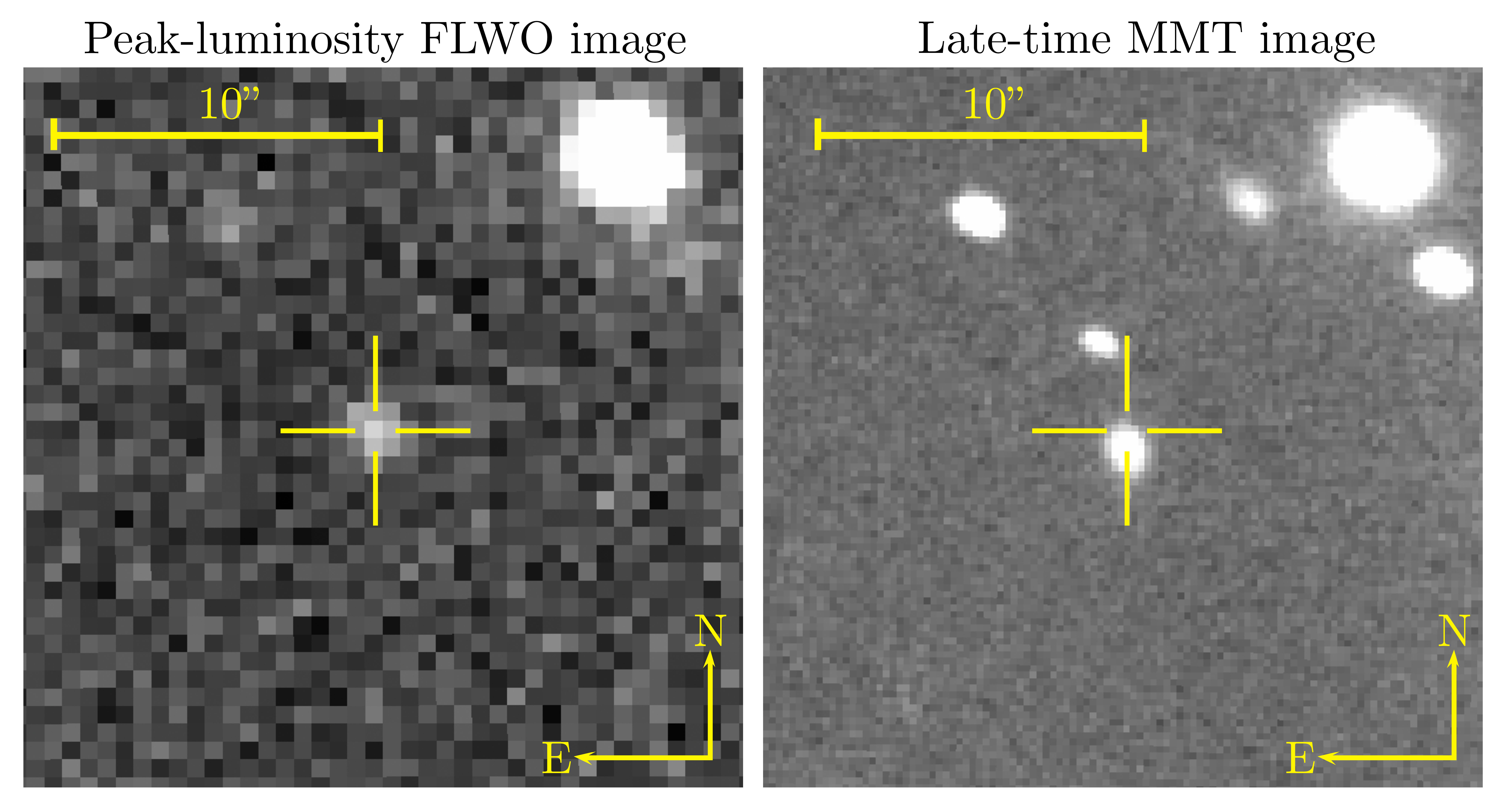}
		\caption{Optical $i$-band images of the field of SN\,2019stc. \textit{Left}: Our first FLWO image taken at MJD = 58818 (phase = 22 days). \textit{Right}: Our last MMT image taken at MJD = 59174 (phase = 340 days) that shows a contribution from only the host galaxy. We see the SN is located on the outskirts of the host galaxy. \label{fig:image}}
	\end{center}
\end{figure}

\subsection{Discovery}

SN\,2019stc was first detected as a transient by the Zwicky Transient Facility (ZTF; \citealt{Bellm19}) on 2019 September 30 at R.A.=${\rm 06^h54^m23^s.10}$, decl.=$+17^\circ29'31''.35$ (J2000) with a magnitude of $m_r = 19.95$ mag and designated ZTF19acbonaa. We triggered follow-up spectroscopic observations of the transient as part of our {\it Finding Luminous and Exotic Extragalactic Transients} (FLEET) observational program designed to find SLSNe \citep{Gomez20}.

\begin{figure*}[t!]
	\begin{center}
		\includegraphics[width=0.9\textwidth]{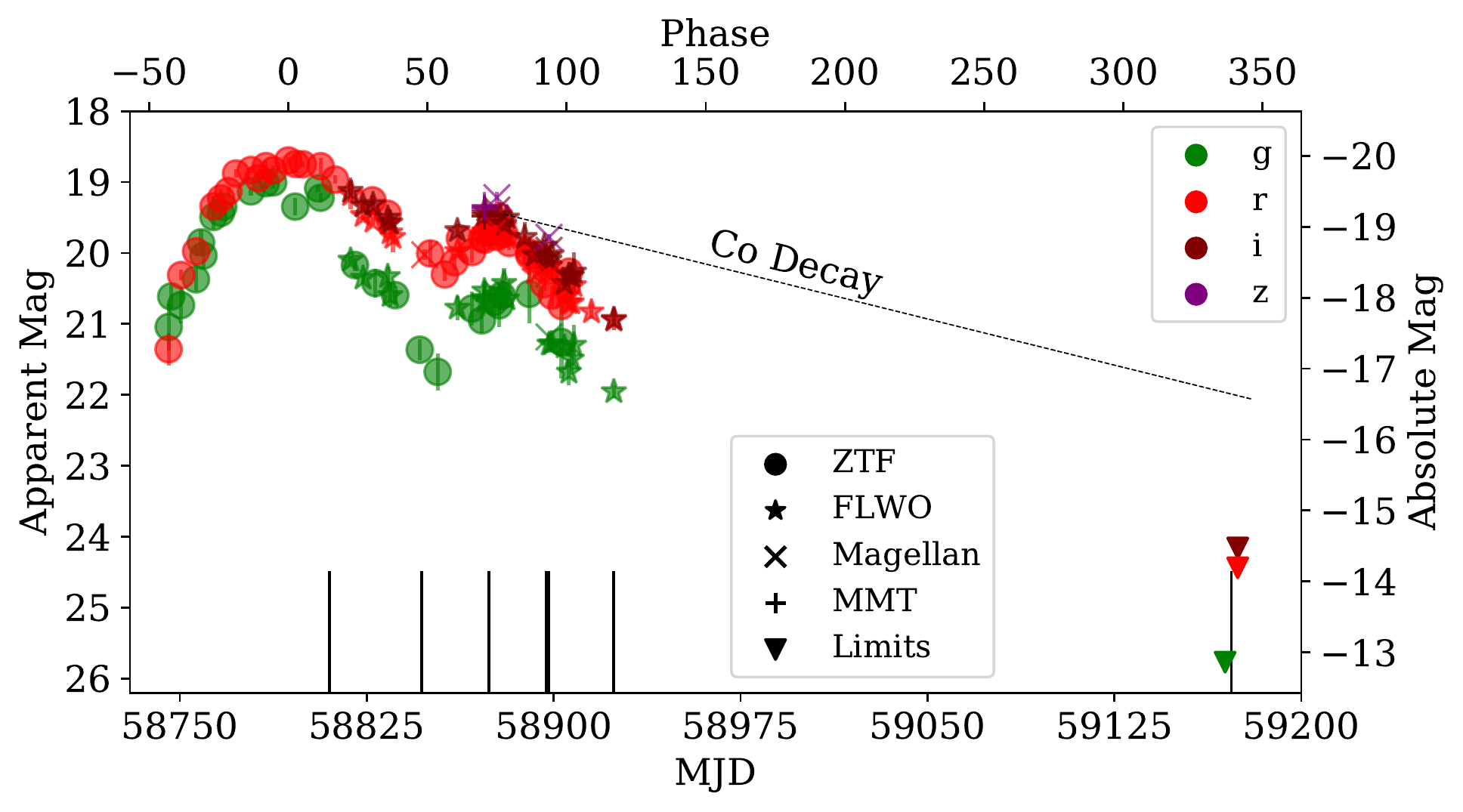}
		\caption{Optical light curves of SN\,2019stc in the $griz$ bands. Magnitudes are in the AB system and not corrected for Galactic extinction. Absolute magnitudes have an additional cosmological K-correction applied. The vertical lines mark the epochs of spectroscopy. The final upper limits are $3\sigma$ non-detections from deep MMT+Binospec images. \label{fig:lightcurve}}
	\end{center}
\end{figure*}

\subsection{Optical Photometry}

We obtained optical images of SN\,2019stc in the $griz$ filters with three different telescopes: KeplerCam on the 1.2-m telescope at Fred Lawrence Whipple Observatory (FLWO); the Low Dispersion Survey Spectrograph (LDSS3c; \citealt{stevenson16}) on the Magellan Clay 6.5-m telescope at Las Campanas Observatory, and Binospec \citep{Fabricant19} on the MMT 6.5-m telescope. We reduced the images using standard IRAF\footnote{\label{ref:IRAF}IRAF is written and supported by the National Optical Astronomy Observatories, operated by the Association of Universities for Research in Astronomy, Inc. under cooperative agreement with the National Science Foundation.} routines, and performed photometry with the {\tt daophot} package.

Instrumental magnitudes were measured by modeling the point-spread function (PSF) of each image using field stars and subtracting the model PSF from the target (Figure~\ref{fig:image}). To separate the flux of SN\,2019stc from that of its host galaxy we perform image subtraction on each image with {\tt HOTPANTS} \citep{Becker15}, using archival PS1/$3\pi$ images as reference templates. We estimate individual zero-points of each image by measuring the magnitudes of field stars and comparing to photometric AB magnitudes from the PS1/$3\pi$ catalog \citep{Chambers18}. The uncertainties reported here are the combination of the photometric uncertainty and the uncertainty in the zero-point determination.

We include additional photometry from ZTF images. In order to recover any detections of SN\,2019stc not reported by the automated ZTF pipeline we downloaded the original ZTF images from the NASA/IPAC Infrared Science Archive\footnote{\url{https://irsa.ipac.caltech.edu/Missions/ztf.html}}. We performed photometry on these images in the same way as the FLWO images described above. All the photometry used for this work is provided in machine readable format in the online version of this paper and on the Open Supernova Catalog \footnote{\label{ref:osc}\url{https://sne.space/}} \citep{guillochon17}. All data are provided before correcting for Galactic extinction and calibrated to PS1 AB magnitudes.

The full multi-color light curves of SN\,2019stc are shown in Figure~\ref{fig:lightcurve}, where we define phase 0 to be the date of the brightest $r$-band magnitude, ${\rm MJD} = 58793.5$. We use this reference date throughout the paper, with all days quoted in the rest frame unless otherwise noted. Absolute magnitudes have an additional cosmological K-correction applied of $+2.5\log\left(1 + z\right)$. All of the photometry is corrected for Galactic extinction using $A_V = 3.1$ and $E(B-V) = 0.084$ mag, according to the dust maps from \cite{Schlafly11}. We use the \citet{Barbary16} implementation of the \citet{Cardelli89} extinction law to correct the photometry. Based on the expectation from case B recombination, we determine that host galaxy extinction corrections are not necessary (described further in \S\ref{sec:host}).

Our final epoch of imaging, in $gri$, was obtained on 2019 November 11 with Binospec on MMT. We do not detect SN\,2019stc down to a magnitude of $r> 24.2$, $g > 25.4$, and $i > 24.0$. These are $3\sigma$ upper limits obtained from forced photometry at the location of SN\,2019stc on difference images with subtracted PS1/$3\pi$ reference templates.

\begin{deluxetable*}{cccccccc}
    \tablecaption{Log of Optical Spectroscopic Observations \label{tab:spectroscopy}}
    \tablewidth{0pt}
    \tablehead{
        \colhead{UT Date} & \colhead{MJD} & \colhead{Phase} & \colhead{Exp.~Time} & \colhead{Airmass} & Wavelength Range & \colhead{Instrument} & \colhead{Grating} \\
        &             & \colhead{(d)}   & \colhead{(s)}           &                   &     (\AA)           &                & (lines/mm)                         
    }
    \startdata
    2019 Nov 23 & 58810.0 & +15  & 1200   & 1.2  & 3820$-$9210 & Binospec     & 270 \\
    2019 Dec 30 & 58847.0 & +48  & 2400   & 1.6  & 4210$-$8890 & IMACS        & 300 \\
    2020 Jan 26 & 58874.0 & +72  & 2700   & 1.2  & 3820$-$9210 & Binospec     & 270 \\
    2020 Feb 18 & 58897.0 & +93  & 1800   & 1.1  & 3320$-$8540 & Blue Channel & 300 \\
    2020 Feb 19 & 58898.0 & +94  & 1500   & 1.5  & 4000$-$9270 & LDSS3c       & 400 \\
    2020 Mar 16 & 58924.0 & +117 & 2400   & 1.0  & 3820$-$9210 & Binospec     & 270 \\
    2020 Nov 19 & 59172.0 & +339 & 4800   & 1.6  & 4800$-$9790 & LDSS3c       & 400 \\
    \enddata
    \tablecomments{All observations were taken with a 1$''$ slit width.}
\end{deluxetable*}

\subsection{Optical Spectroscopy}

\begin{figure}[t!]
	\begin{center}
		\includegraphics[width=\columnwidth]{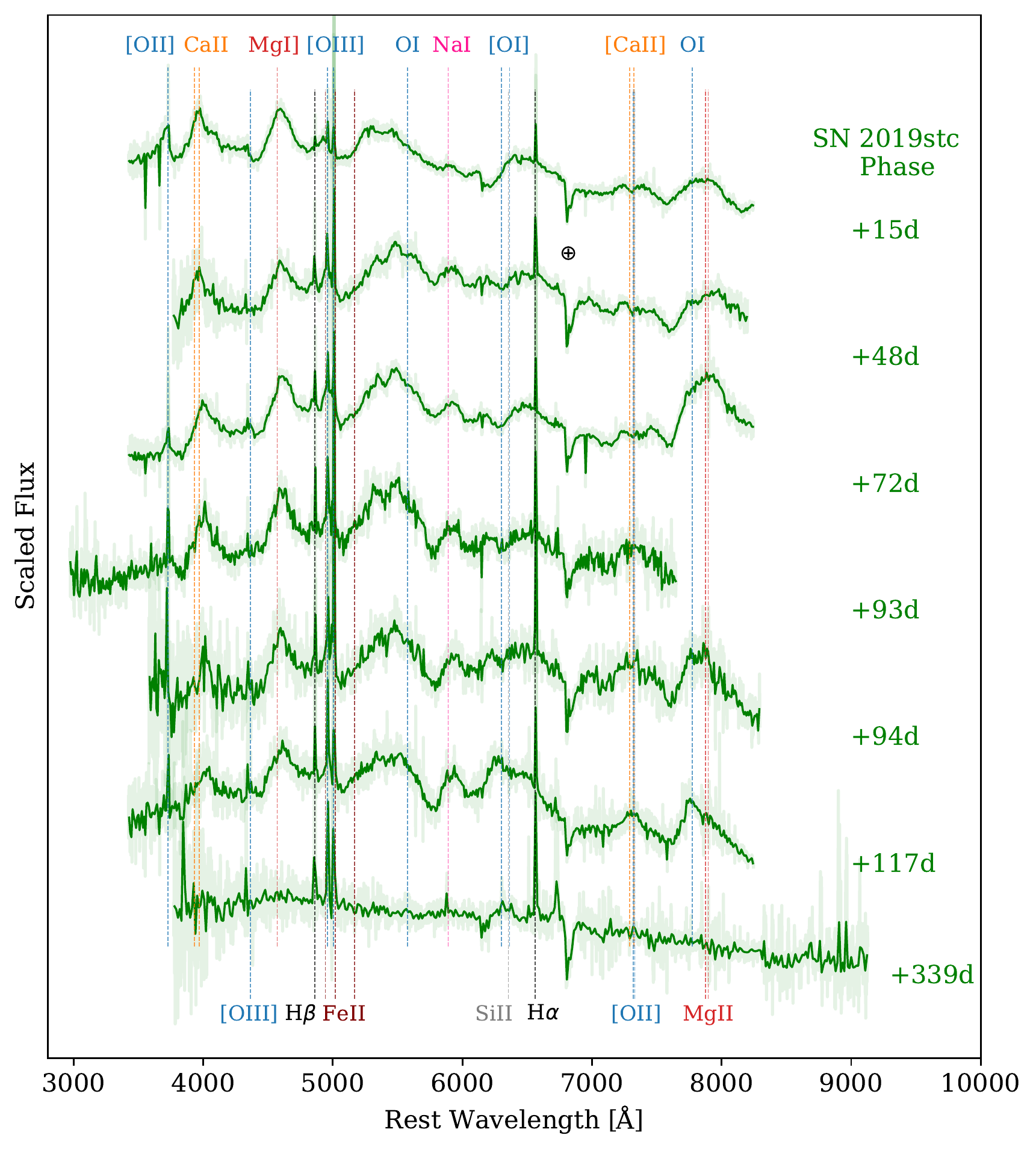}
		\caption{Optical spectra of SN\,2019stc, corrected for Galactic extinction and shifted to the rest-frame of SN\,2019stc using $z=0.117$. We binned the spectra for clarity and show the unbinned spectra as shaded regions. Relevant spectral lines are marked. The last spectrum at a phase of 339 days is dominated by host galaxy emission. \label{fig:spectra}}
	\end{center}
\end{figure}

We obtained seven epochs of low-resolution optical spectroscopy covering phases from 15 to 340 days. We used the LDSS3c Spectrograph \citep{stevenson16} and Inamori-Magellan Areal Camera and Spectrograph (IMACS; \citealt{dressler11}) on the Magellan 6.5-m telescopes and the Blue Channel \citep{schmidt89} and Binospec \citep{fabricant03} spectrographs on the MMT 6.5-m telescope. All the spectra were obtained with the slit oriented along the parallactic angle. Details of the spectroscopic observations are provided in Table~\ref{tab:spectroscopy}.

We reduced the spectra using standard IRAF$^{\ref{ref:IRAF}}$ routines using the {\tt twodspec} package. The spectra were bias-subtracted and flat-fielded, the sky background was modeled and subtracted from each image, and the one-dimensional spectra were optimally extracted, weighing by the inverse variance of the data. A wavelength calibration was applied using an arc lamp spectrum taken near the time of each science image. Relative flux calibration was applied to each spectrum using a standard star taken close to the time of observation. The spectra were then calibrated to absolute flux using our $gri$ photometry as a reference, applying a normalization constant to each spectrum to match the expected flux from the photometry using the {\tt PYPHOT} Python package. Lastly, the spectra were corrected for Galactic extinction using \mbox{$E(B-V) = 0.084$} and transformed to the rest frame of SN\,2019stc using $z = 0.117$. All the spectroscopy collected for this work is made available on WISeREP\footnote{\url{https://wiserep.weizmann.ac.il/}} \citep{Yaron12} and the Open Supernova Catalog $^{\ref{ref:osc}}$ \citep{guillochon17}, after flux calibrating but before correcting for Galactic extinction and redshift. The spectra are shown in Figure~\ref{fig:spectra}.

\section{Light Curves}
\label{sec:analysis} 

\begin{figure}[t!]
	\begin{center}
		\includegraphics[width=\columnwidth]{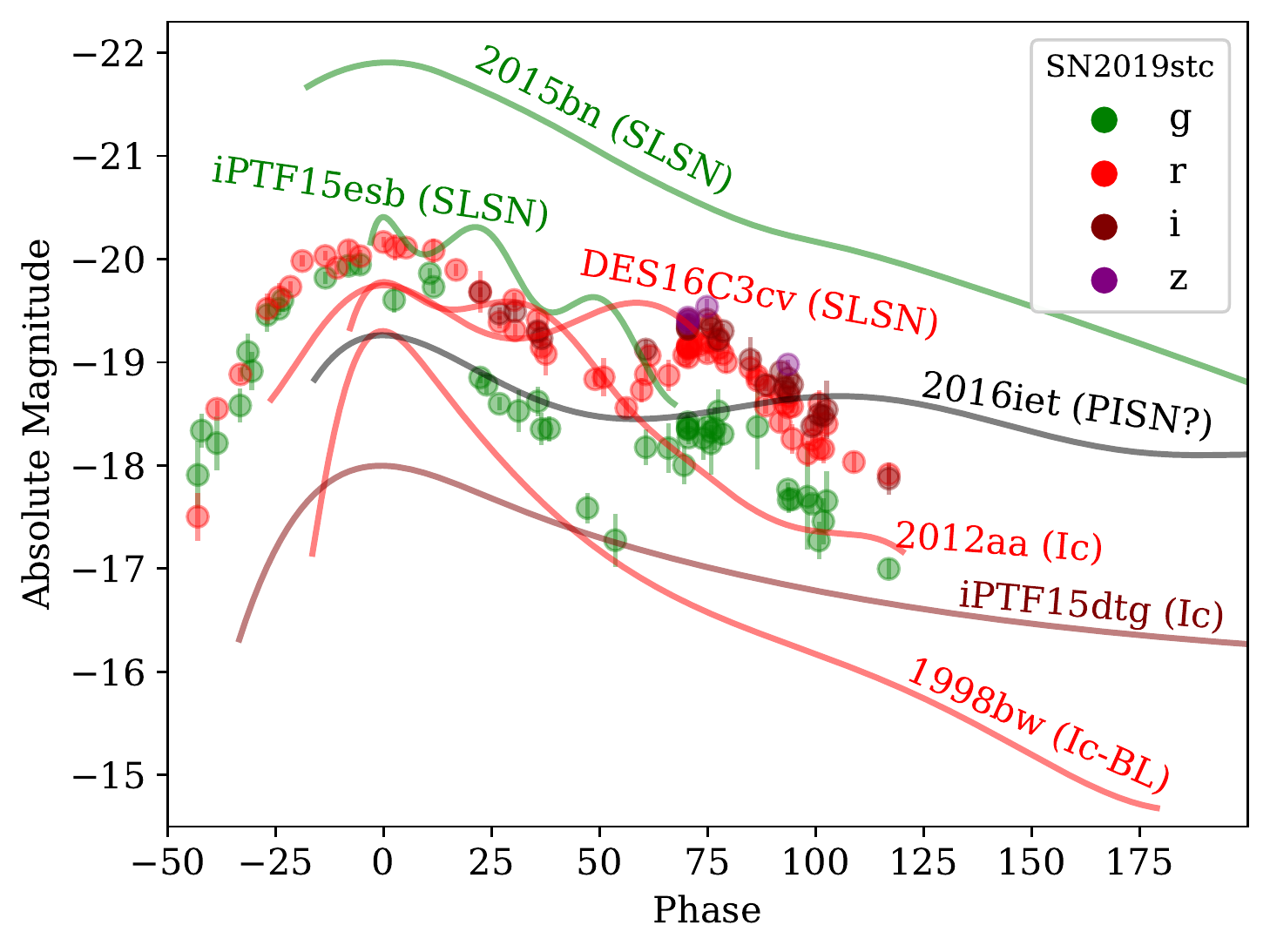}
		\caption{Light curve of SN\,2019stc compared to several other luminous and/or double-peaked hydrogen-poor SNe: SN\,2015bn \citep{Nicholl16_15bn, Nicholl16_15bnnebular,Nicholl18_15bn}, iPTF15esb \citep{yan17}, SN\,2016iet \citep{Gomez19}, SN\,2012aa \citep{Roy16}, iPTF15dtg \citep{Taddia19_latetime}, DES16C3cv \cite{Angus19}, and SN\,1998bw \citep{galama98}. The comparison light curves have been smoothed for clarity using a 1-D spline and shifted in phase to match the first peak of SN\,2019stc. For SN\,2016iet we smooth a combination of G, C, and $i$ bands given its sparse multi-band coverage at early times. None of the comparison SNe show a double-peak structure as pronounced as SN\,2019stc. \label{fig:comparisons}}
	\end{center}
\end{figure}

The light curve of SN\,2019stc exhibits a peculiar and pronounced double-peaked structure, clearly seen in all available filters (Figure~\ref{fig:lightcurve}). The first peak reaches $M_r = -20.0$ at 58793.5 MJD, with a rise time of $\sim 40$ days. Subsequently, the light curve declines at a rate of $0.033\pm 0.004$ mag day$^{-1}$. $\sim 53$ rest days after the first peak, the light curve begins to re-brighten and produces a second peak with $M_r = -19.2$ (at 58882 MJD), 79 rest-frame days after the first peak. After the second peak, the light curve declines at a rate of $0.036\pm 0.003$ mag day$^{-1}$. This decline rate is much too fast to be consistent with Co decay's rate of $\approx 0.01$ mag day$^{-1}$, suggesting that the second peak is not due to radioactive decay.

In Figure~\ref{fig:comparisons} we compare the light curve of SN\,2019stc to those of both normal and long-lived stripped SNe, with a specific focus on luminous events and other events that have showed a double peaked structure. The first peak of SN\,2019stc is more luminous and broader than even the broad-lined Type Ic SN\,1998bw \citep{galama98}, much less bright than the archetype SLSN\,2015bn \citep{Nicholl16_15bn}, and significantly brighter than normal SNe Ic (e.g., iPTF15dtg \citealt{Taddia19_latetime}). Other events such as iPTF15esb \citep{yan17} or SN\,2012aa \citep{Roy16} exhibit undulations in their light curves, but all with a much lower amplitude and on much shorter timescales than SN\,2019stc. Spectroscopically, iPTF15esb is a SLSN that showed late-time H$\alpha$ emission, while SN\,2012aa has a spectral sequence that shares similarities that place is between the SLSNe and SNe Ic population. A similar light curve structure was also seen in the SLSN DES16C3cv, where the two peaks are separated by $\approx 60$ days \citep{Angus19}. Additionally, we show SN\,2016iet, a hydrogen and helium free pulsational pair-instability SN (PPISN) candidate in which the two peaks were separated by 100 days and are comparable in luminosity to SN\,2019stc \citep{Gomez19}. However, we note that the peaks in DES16C3cv and SN\,2016iet are less pronounced than in SN\,2019stc. Therefore, we conclude that SN\,2019stc is unique among the hydrogen/helium-poor SN class in terms of its light curve properties.

\begin{figure}
	\begin{center}
		\includegraphics[width=0.8\columnwidth]{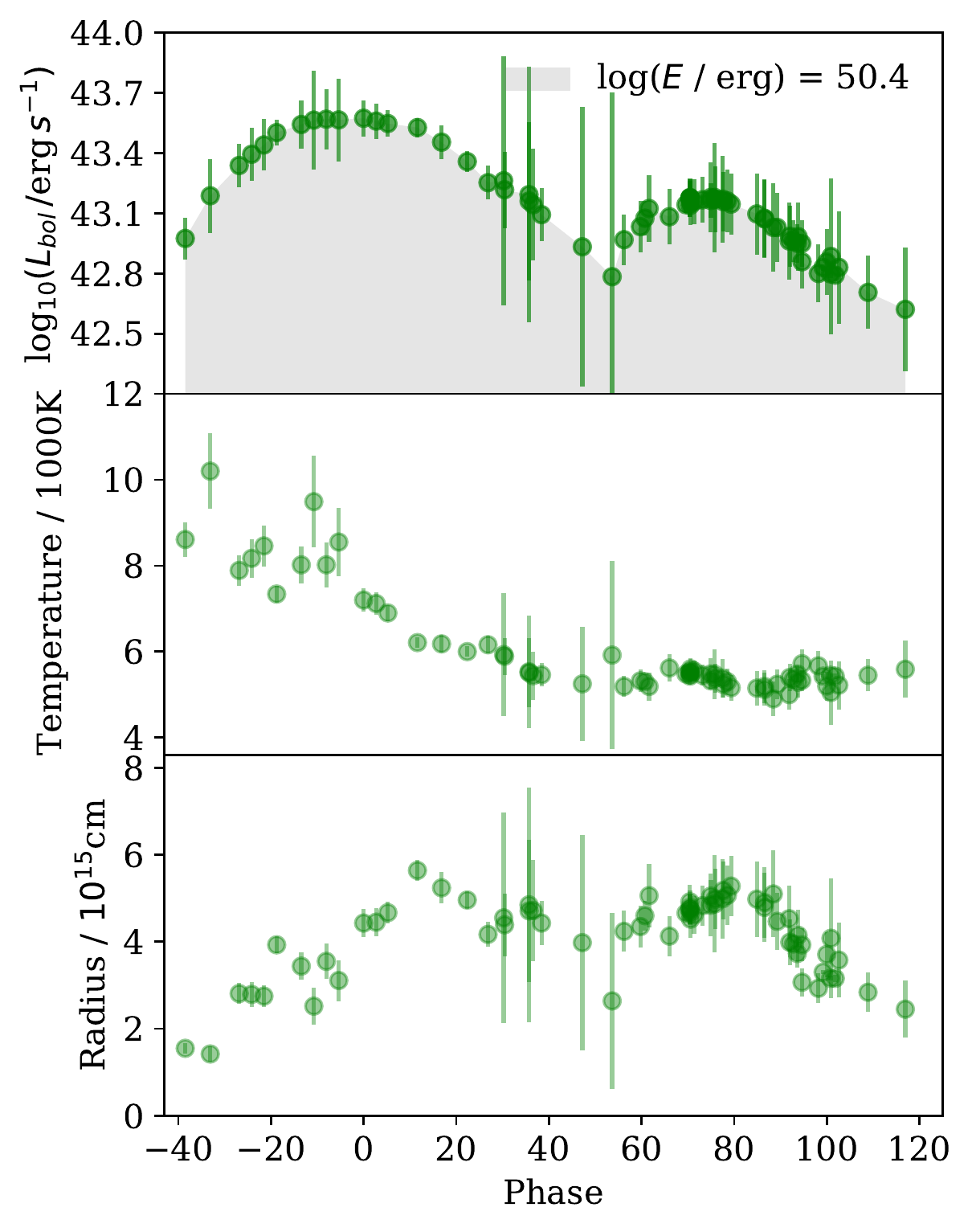}
		\caption{Time evolution of the bolometric light curve ({\it Top}), blackbody temperature ({\it Middle}), and photospheric radius ({\it Bottom}) of SN\,2019stc. The shaded gray region indicates the area over which we determine the total radiated energy, which we find to be $E_{\rm rad}\approx 2.5\times 10^{50}$ erg. The increase in photospheric radius during the second peak is possibly indicative of CSM interaction and the formation of a new photosphere in the interaction region. \label{fig:bolometric}}
	\end{center}
\end{figure}

\begin{figure*}
	\begin{center}
		\includegraphics[width=\textwidth]{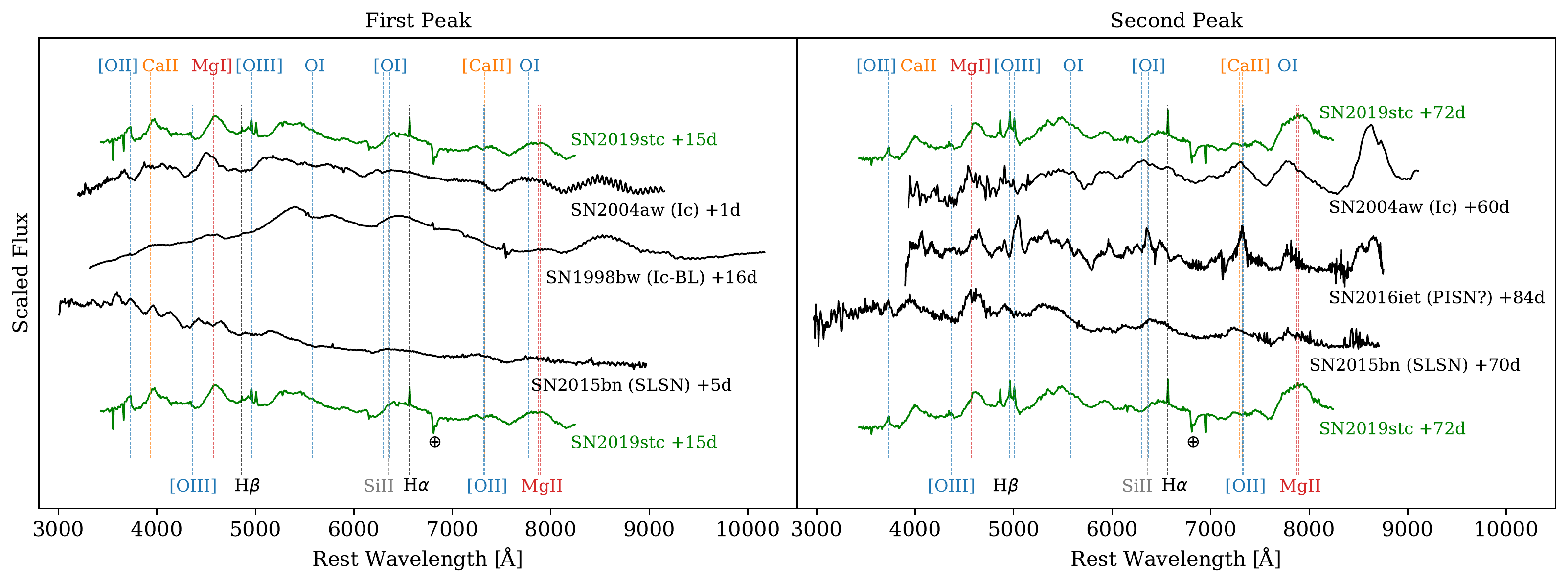}
		\caption{Spectra of SN\,2019stc (green) during the first ({\it Left}) and second ({\it Right}) peaks, compared to other representative stripped envelope SNe (black): SN\,1998bw \citep{Patat01}, SN\,2004aw \citep{Modjaz14}, SN\,2016iet \citep{Gomez19}, and SN\,2015bn \citep{Nicholl16_15bn}). The vertical lines mark the locations of dominant spectral features and $\oplus$ marks telluric features. The SN\,2019stc spectra has been binned for clarity. We find that SN\,2019stc most closely matches the normal SN Ic SN\,2004aw, and is clearly distinct from the other comparison SNe. \label{fig:spectra_comparison}}
	\end{center}
\end{figure*}

We calculate the bolometric light curve of SN\,2019stc, its photospheric radius, and temperature evolution by fitting the spectral energy distribution (SED) of each epoch with a blackbody function using the {\tt Superbol} code \citep{Nicholl18_superbol}. We further extrapolate the blackbody SED to account for the missing coverage in the UV and NIR. The resulting bolometric light curve, photospheric radius, and temperature evolution are shown in Figure~\ref{fig:bolometric}. The error bars represent the uncertainties in the measurement of luminosity, temperature, and radius; which are affected by a combination of the photometric error bars, the uncertainty in the light curve interpolation, and the blackbody fits. The error bars for the epochs between $\sim 30$ to $\sim 50$ days are particularly affected due to the uncertainty in the interpolation when the light curve is turning over. To mittage this issue and due to the fact we only have photometry in two bands during this epoch, we include one synthetic $i$-band data point to the {\tt Superbol} interpolation at MJD = 58847.0, obtained from a convolution of the corresponding spectrum. The total integrated energy in the observed epochs between a phase of $-38$ and 117 days is $E_{\rm rad}\approx 2.5\times 10^{50}$ erg. This is over an order of magnitude larger than for normal SNe Ic \citep{Prentice16}.

The blackbody temperature during the first peak is about 8500 K, which then steadily decreases to about 5500 K by the time the second peak begins and stays at that value throughout the remainder of the light curve. The photospheric radius initially increases through the first peak to a maximum value of about $5.5\times10^{15}$ cm, then recedes to about $4\times 10^{15}$ cm, increases again through the second peak (to about $5.5\times10^{15}$ cm), and finally declines steadily following the second peak. We note that the increase in photospheric radius during the second peak could be a signature of CSM interaction, as a new photosphere would form in the interaction region.

\section{Spectral Features}
\label{sec:spectra}

Our optical spectra of SN\,2019stc span 15 to 339 days after peak (Figure~\ref{fig:spectra}). The spectra are typical of a normal SN Ic, with no evidence of hydrogen or helium lines and a downturn in the flux blueward of about 5000 \AA. This is different from most SLSNe, which exhibit a significantly bluer continuum (e.g., \citealt{Howell13, Nicholl16_15bn, Quimby18}), although some do show a similar downturn below $\sim 5000$ \AA\ (e.g., SN\,2017dwh; \citealt{Blanchard19}). We do not see a dramatic evolution in the spectra from 15 days after peak to 72 days after peak, other than a redder color. The spectra are dominated by broad lines of Mg, Ca, Fe, and O. We fit the absorption component present in the \ion{O}{1} $\lambda7774$, \ion{Na}{1} $\lambda5890$, \ion{Mg}{1}] $\lambda4571$, and \ion{Si}{2} $\lambda6355$ lines with a Gaussian profile and use the center of these to estimate the velocity of the ejecta. We obtain an average ejecta velocity by measuring the center of the absorption component of the \ion{O}{1} $\lambda7774$ lines in our spectra and obtain a value of $7800\pm 1900$ km s$^{-1}$, roughly constant and comparable to the velocities of SNe Ic near peak \citep{Taddia19_broadlined}.

In Figure~\ref{fig:spectra_comparison} we compare the spectra of SN\,2019stc during the first and second peaks to those of several stripped SNe: SN\,1998bw (Ic-BL; \citealt{Patat01}), SN\,2004aw (Ic; \citealt{Modjaz14}), SN\,2016iet (PISN candidate; \citealt{Gomez19}), and SN\,2015bn (SLSN; \citealt{Nicholl16_15bn}). During the first peak the spectral features in SN\,2019stc are significantly narrower than in the broad-lined SN\,1998bw (which has $V\approx 30,000$ km s$^{-1}$), and slower than in the normal Ic SN\,2004aw, which had a velocity of $V_{\rm ej}\approx 11,100$ km s$^{-1}$ near peak \citep{Taubenberger06, Mazzali17}. The spectra of SN\,2019stc are significantly redder at $\lesssim 5000$ \AA\ than that of the SLSN 2015bn and lack the characteristic oxygen features near $\sim 4000$ \AA\ found in SLSNe. During the second peak, SN\,2019stc still closely resembles SN\,2004aw, and is still redder than SN\,2015bn. It also appears distinct from SN\,2016iet, which partially resembles a SN Ic, but has a bluer continuum and narrower spectral features than SN\,2019stc. We therefore conclude that spectroscopically SN\,2019stc most closely matches a normal SN Ic during both the first and second peak.

\section{Light Curve Modeling} 
\label{sec:modeling} 

We model the light curves of SN\,2019stc using the Modular Open-Source Fitter for Transients ({\tt MOSFiT}) Python package, a flexible code that uses a Markov chain Monte Carlo (MCMC) implementation to fit the light curves of transients using a variety of different power sources \citep{guillochon18}. We run each MCMC using an implementation of the {\tt emcee} sampler \citep{foreman13}. We test for convergence by ensuring that the models reach a potential scale reduction factor of $<1.2$ \citep{gelman92}, which corresponds to about $35,000$ steps with 150 walkers. The definitions of all the parameters used in this section are listed in Table~\ref{tab:parameters}. The uncertainties presented here represent only the statistical errors on the fits.

An individual power source in MOSFiT, such as a magnetar or radioactive decay, can only reproduce a single peaked light curve. Therefore, we first investigate the origin of the first peak and include only the data up to MJD = 58853 (Phase = 53 days), before the emergence of the second peak. We explore the origin of the second peak in \S\ref{sec:second}.

\begin{deluxetable}{ll}
	\tablecaption{MOSFiT Parameter Definitions \label{tab:parameters}}
	\tablehead{\colhead{Parameter} & \colhead{Definition}}
	\startdata
	$M_{\text{ej}}$        & Ejecta mass  \\
	$M_{\text{Ni}}$        & Radioactive nickel mass  \\
	$f_{\text{Ni}}$        & Nickel mass as a fraction of the  ejecta mass  \\
	$v_{\text{ej}}$        & Ejecta velocity  \\
	$E_{k}$                & Ejecta kinetic energy   \\
	$R_{0}$                & CSM inner radius  \\
	$\rho_{0}$             & CSM density at $R_0$  \\
	$M_{\text{NS}}$        & Neutron star mass   \\
	$P_{\text{spin}}$      & Magnetar spin   \\
	$B_{\perp}$            & Magnetar magnetic field strength \\
	$\theta_{\text{BP}}$   & Angle of the dipole moment \\
	$t_{\text{exp}}$       & Explosion time relative to first data point  \\
	$T_{\text{min}}$       & Photosphere temperature floor  \\
	$\lambda_{\text{cut}}$ & Flux below this wavelength is suppressed \\
	$n_{H,\text{host}}$    & Column density in the host galaxy \\
	$A_{V, \text{host}}$   & Extinction in the host galaxy   \\
	$\kappa$               & Optical opacity \\
	$\kappa_{\gamma}$      & Gamma-ray opacity  \\
	$\sigma$               & Uncertainty required for $\chi^2_r=1$ \\
	\enddata
\end{deluxetable}

\begin{deluxetable}{cccc}
    \tablecaption{Best-fit Parameters for the Radioactive Decay Model \label{tab:radioactive}}
    \tablehead{\colhead{Parameter} & \colhead{Prior}  & \colhead{Posterior}  & \colhead{Units}}
    \startdata
    $M_{\text{ej}}                $ & $[0.1, 100]$      &    $ 10.3^{+1.9}_{-1.6}     $     &  M$_\odot$      \\
    $f_{\text{Ni}}                $ & $[0.0, 1]$        &    $ 0.31^{+0.06}_{-0.05}   $     &                 \\
    $M_{\text{Ni}}^{\dagger}      $ &                   &    $ 3.2 \pm 0.2            $     &  M$_\odot$      \\
    $V_{\text{ej}}                $ & $7800\pm1900$     &    $ 8000^{+560}_{-540}     $     &  km s$^{-1}$    \\
    $E_{k}^{\dagger}              $ &                   &    $ 3.9^{+1.3}_{-1.0}      $     &  $10^{51}$ erg  \\
    $t_{\text{exp}}               $ & $[0, 50]$         &    $ 7.3 \pm 0.7            $     &  days           \\
    $T_{\text{min}}               $ & $[3000, 10^5] $   &    $ 3900^{+620}_{-600}     $     &  K              \\
    $\log{(n_{H,\text{host}})}    $ & $[16, 23]$        &    $ 18.4 \pm 1.6           $     &  cm$^{-2}$      \\
    $A_{V, \text{host}}^{\dagger} $ &                   &    $ < 0.05                 $     &  mag            \\
    $\log{(\kappa_{\gamma})}      $ & $[-2, 2]$         &    $ -1.96^{+0.05}_{-0.03}  $     &  cm$^2$g$^{-1}$ \\
    $\log{\sigma}                 $ & $[-3, 2]$         &    $ -0.77 \pm 0.06         $     &                 \\
    \enddata
    \tablecomments{Best model parameters, prior ranges, and 1$\sigma$ uncertainties for the model realizations shown in Figure~\ref{fig:radioactive}. See Table~\ref{tab:parameters} for parameter definitions.}
    \tablenotetext{\dagger}{These parameters are calculated using the posterior distribution samples of the fitted parameters.}
\end{deluxetable}

\subsection{Radioactive Decay}
\label{sec:radioactive}

Motivated by the SN Ic spectrum of SN\,2019stc, we first investigate a light curve model with heating from nickel-cobalt radioactive decay \citep{Arnett82} where the input luminosity is given by:
\begin{equation}
L_\gamma = M_{\rm Ni}\left[\epsilon_{\rm Ni}e^{-t/\tau_{\rm Ni}}+\epsilon_0 e^{-t/\tau_{\rm Co}}\right].
\end{equation}
Here $t$ is rest-frame time; $M_{\rm Ni}$ is the initial mass of $^{56}$Ni; $\tau_{\rm Ni}=8.8$ days and $\tau_{\rm Co}=111.3$ days are the half-lives of $^{56}$Ni and $^{56}$Co, respectively \citep{nadyozhin94}; \mbox{$\epsilon_{\rm Ni} = 6.45\times10^{43}$ erg s$^{-1}$ M$_\odot^{-1}$} is the heating rate for $^{56}$Ni; and $\epsilon_0=1.45\times10^{43}$ erg s$^{-1}$ M$_\odot^{-1}$ is the effective heating rate for $^{56}$Co, which is a function of $\tau_{\rm Ni}$ and $\tau_{\rm Co}$. We assume an opacity of $\kappa = 0.07$, determined to be a suitable value for SNe Ic \citep{Taddia18}.

\begin{figure}[t!]
	\begin{center}
		\includegraphics[width=\columnwidth]{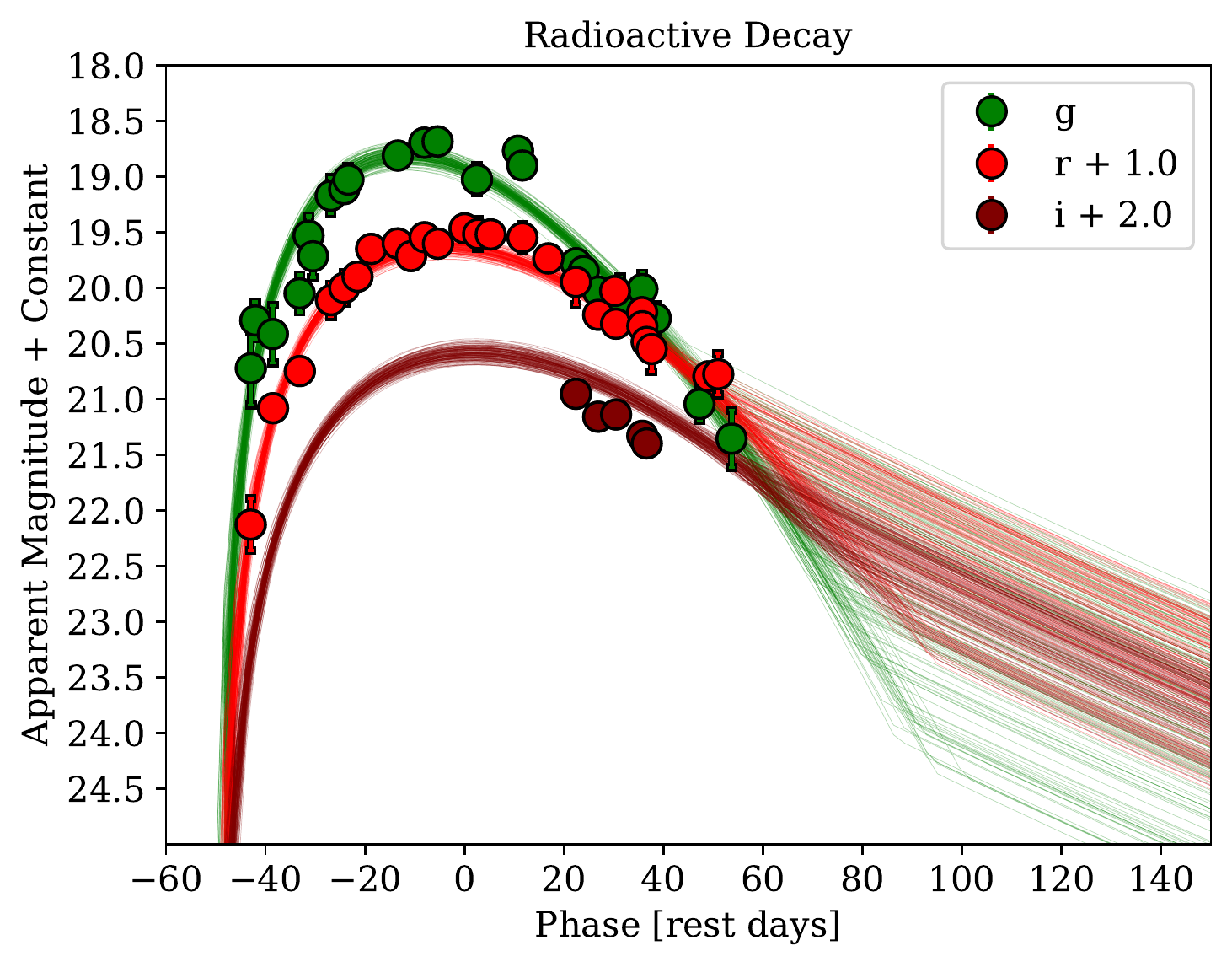}
		\caption{First peak of the light curve of SN\,2019stc with the best fit {\tt MOSFiT} radioactive decay model described in \S\ref{sec:radioactive}. Corrected for Galactic extinction. \label{fig:radioactive}}
	\end{center}
\end{figure}

\begin{figure}[t!]
	\begin{center}
		\includegraphics[width=\columnwidth]{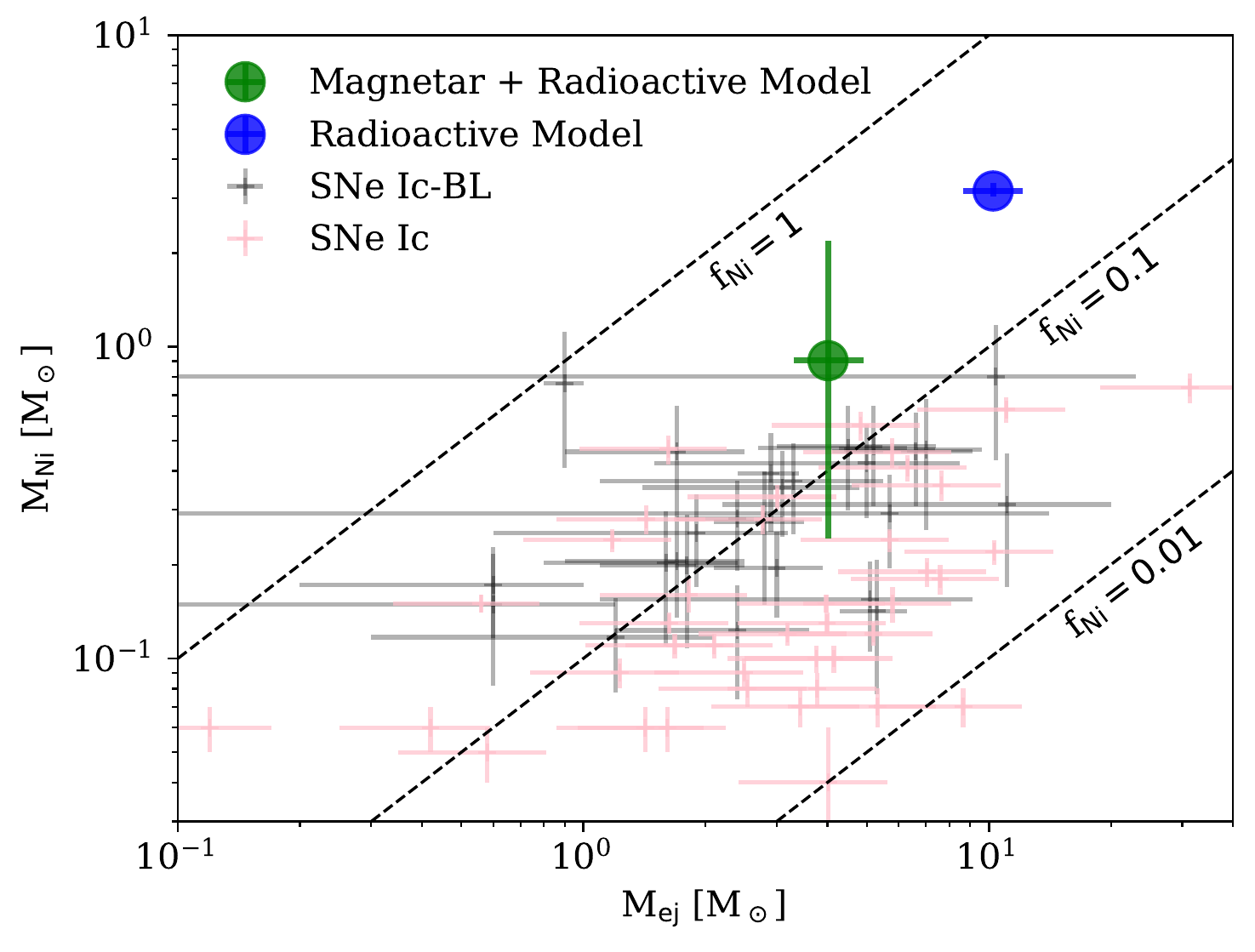}
		\caption{The $^{56}$Ni mass versus ejecta mass inferred from the radioactive decay (blue) and the magnetar plus radioactive decay (green) models of SN\,2019stc. Also shown for comparison are SNe Ic-BL (grey; \citealt{Taddia19_broadlined}) and SNe Ic (pink; \citealt{Barbarino20}) populations. Representative nickel fractions of 0.01, 0.1, and 1 are shown with dashed lines. The inferred values of $M_{\rm ej}$ and $M_{\rm Ni}$ for the radioactive decay model of SN\,2019stc are well outside of the distribution for the SN Ic and Ic-BL population, while the values from the magnetar plus radioactive decay models are consistent with the SN Ic and Ic-BL population. \label{fig:taddia}}
	\end{center}
\end{figure}

We show the best-fit realizations of this model in Figure~\ref{fig:radioactive} and list the best-fit parameters and associated uncertainties in Table~\ref{tab:radioactive}. We find that the model provides a good fit to the $g$- and $r$-band light curves, but systematically over-predicts the $i$-band data due to a mismatch between the model and data photospheric temperatures.

As expected from the high luminosity and broad light curve, we find large values of $M_{\rm Ni}\approx 3.2$ M$_\odot$ and $M_{\rm ej}\approx 10.3$ M$_\odot$, leading to a high nickel fraction of $f_{\rm Ni}\approx 0.31$. The nickel mass is substantially higher than the typical range for SNe Ic and Ic-BL of $\approx 0.05-0.6$ M$_\odot$  \citep{Drout11,Taddia19_broadlined}. In Figure~\ref{fig:taddia} we compare the results for SN\,2019stc to the SNe Ic sample from \citet{Barbarino20} and the Ic-BL sample from \citet{Taddia19_broadlined}\footnote{We exclude iPTF15eov, which is most likely a SLSN as opposed to a SN Ic-BL.}. We see that the nickel fraction found for SN\,2019stc is much higher than typical SNe Ic, especially in the high ejecta mass regime (Figure~\ref{fig:taddia}). We therefore conclude that while SN\,2019stc most closely resembles SNe Ic in terms of its spectra, it is unlikely to be powered exclusively by radioactive decay.

\begin{deluxetable}{cccc}
    \tablecaption{Best-fit Parameters for the Magnetar Engine + Radioactive Decay Model \label{tab:magnetar}}
    \tablehead{\colhead{Parameter} & \colhead{Prior}  & \colhead{Best-fit}  & \colhead{Units}}
    \startdata
    $M_{\text{ej}}                $ & $[0.1, 100]$       & $ 4.0^{+0.9}_{-0.7}    $ &  M$_\odot$       \\
    $f_{\text{Ni}}                $ & $[0.0, 1]$         & $ 0.2^{+0.3}_{-0.2}    $ &                  \\
    $M_{\text{Ni}}^{\dagger}      $ &                    & $ 0.9^{+1.3}_{-0.7}    $ &  M$_\odot$       \\
    $M_{\text{NS}}                $ & $[1.0, 2.2]$       & $ 1.7 \pm 0.4          $ &  M$_\odot$       \\
    $P_{\text{spin}}              $ & $[0.7, 20]$        & $ 7.2^{+2.7}_{-1.8}    $ &  ms              \\
    $\theta_{\text{PB}}           $ & $[0.0, \pi/2]$     & $ 0.9 \pm 0.4          $ &  rad             \\
    $B_{\perp}                    $ & $\log([0.01, 10])$ & $ 0.9^{+1.0}_{-0.5}    $ &  $10^{14}$ G     \\
    $V_{\text{ej}}                $ & $7800\pm1900$      & $ 6970^{+490}_{-470}   $ &  km s$^{-1}$     \\
    $E_{k}^{\dagger}              $ &                    & $ 1.2^{+0.4}_{-0.3}    $ &  $10^{51}$ erg   \\
    $t_{\text{exp}}               $ & $[0, 50]$          & $ 9.9 \pm 1.5          $ &  days            \\
    $T_{\text{min}}               $ & $[3000, 10^5]$     & $ 4100^{+780}_{-760}   $ &  K               \\
    $\log{(n_{H,\text{host}})}    $ & $[16, 23]$         & $ 18.6 \pm 1.7         $ &  cm$^{-2}$       \\
    $A_{V, \text{host}}^{\dagger} $ &                    & $ < 0.09               $ &  mag             \\
    $\kappa                       $ & $[0.05, 0.2]$      & $ 0.18^{+0.02}_{-0.03} $ &                  \\
    $\log{(\kappa_{\gamma})}      $ & $[-2, 2]$          & $ -1.9 \pm 0.1         $ &  cm$^2$g$^{-1}$  \\
    $\log{\sigma}                 $ & $[-3, 2]$          & $ -0.84 \pm 0.06       $ &                  \\
    \enddata
    \tablecomments{Best model parameters, prior ranges, and 1$\sigma$ uncertainties for the model realizations shown in Figure~\ref{fig:magnetar}. See Table~\ref{tab:parameters} for parameter definitions.}
    \tablenotetext{\dagger}{These parameters are calculated using the posterior distribution samples of the fitted parameters.}
\end{deluxetable}

\begin{figure}[t!]
	\begin{center}
		\includegraphics[width=\columnwidth]{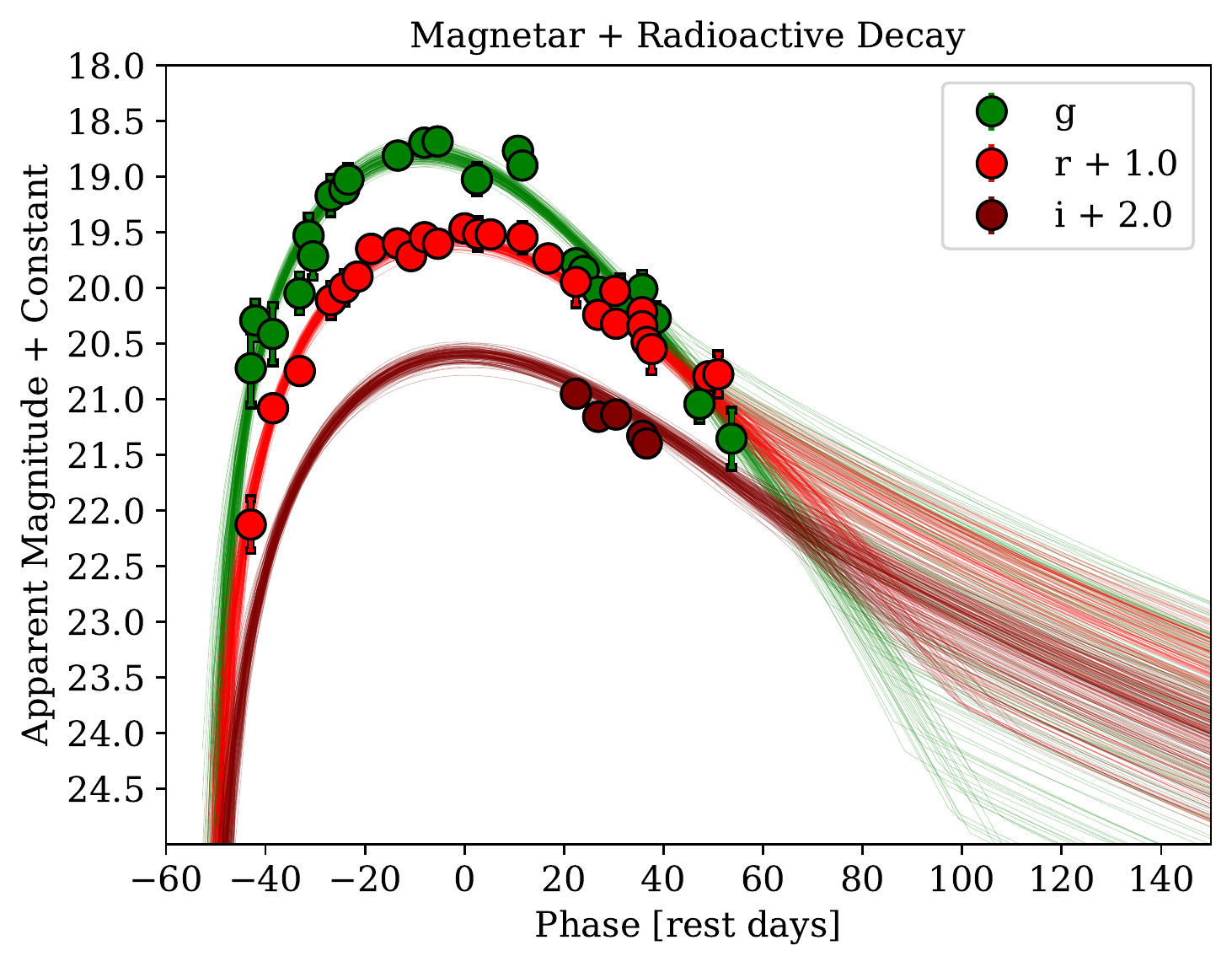}
		\caption{First peak of the light curve of SN\,2019stc with the best fit {\tt MOSFiT} magnetar engine plus radioactive decay model described in \S\ref{sec:magnetar}. Corrected for Galactic extinction. \label{fig:magnetar}}
	\end{center}
\end{figure}

\subsection{Magnetar Engine + Radioactive Decay}
\label{sec:magnetar}

\begin{figure}[t!]
	\begin{center}
		\includegraphics[width=\columnwidth]{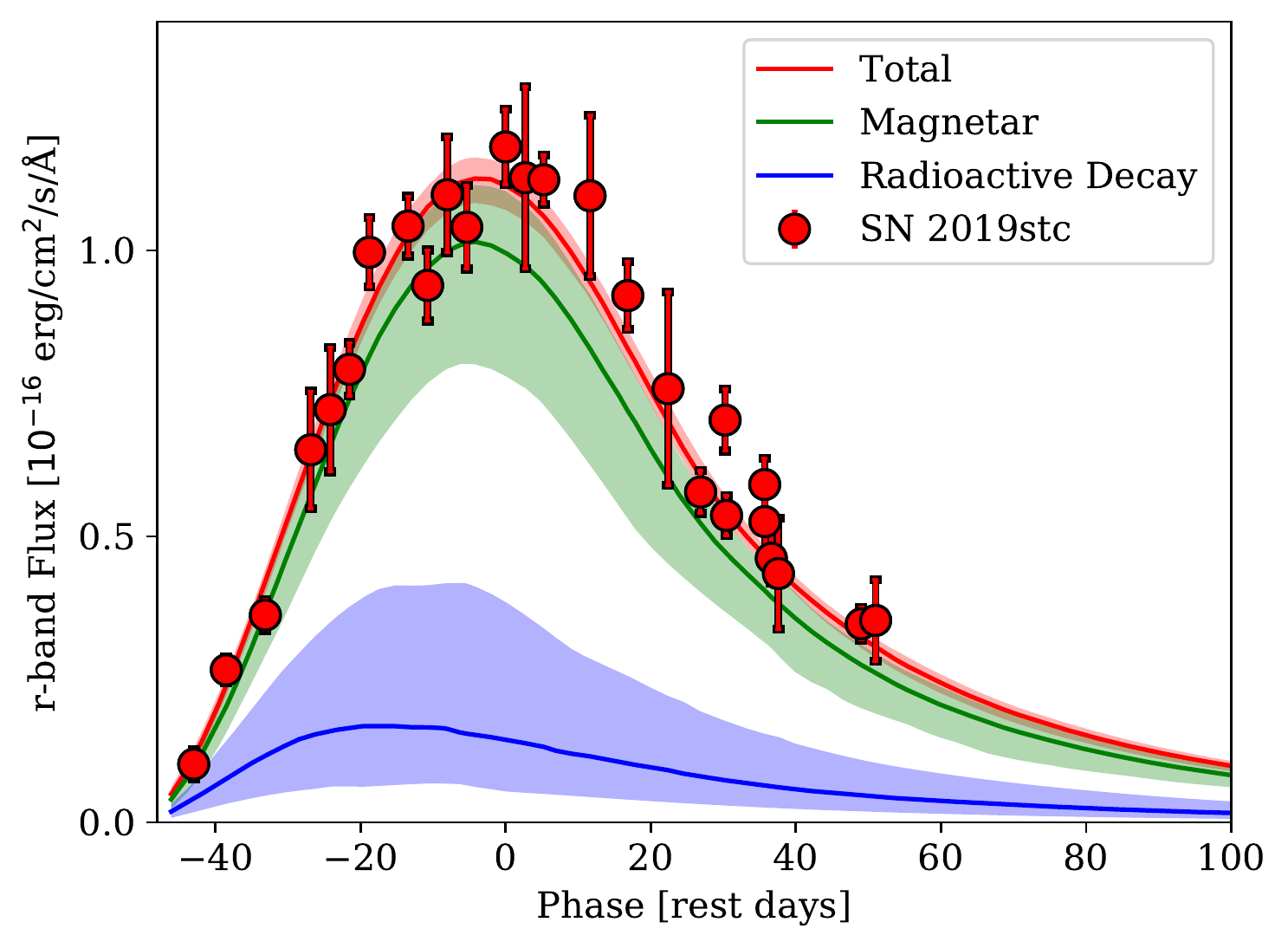}
		\caption{Best fit model realizations (lines) and their $1\sigma$ confidence intervals (shaded regions) for the magnetar engine plus radioactive decay model (red). We decompose the contribution from the magnetar engine (green) and radioactive decay (blue) components. We find that the magnetar engine clearly dominates the energy input. \label{fig:mosfit}}
	\end{center}
\end{figure}

We next explore a model with energy input from both radioactive decay (motivated by the spectral match to a SN Ic) and a magnetar central engine (motivated by the broad and luminous first peak). The MOSFiT setup for the magnetar model \citep{Kasen10,Woosley10} is described in detail in \citet{Nicholl17}. We show the best-fit realizations of this model in Figure~\ref{fig:magnetar}, and list the best-fit parameters and associated uncertainties in Table~\ref{tab:magnetar}. We find an ejecta mass of $M_{\rm ej}\approx 4$ M$_\odot$ and a nickel fraction that is consistent with 0 but extends to the typical values for SNe Ic (Figure~\ref{fig:taddia}). The best fit parameters of the magnetar engine are $P_{\rm spin}\approx 7.2$ ms and $B_{\perp}\approx 1\times10^{14}$ G. In Figure~\ref{fig:mosfit} we break down the model into its two components and find that the magnetar engine dominates the energy input, accounting for $89\pm8$\%, while the radioactive decay component contributes only $11\pm4$\% of the total energy.

\begin{figure}
	\begin{center}
		\includegraphics[width=0.7\columnwidth]{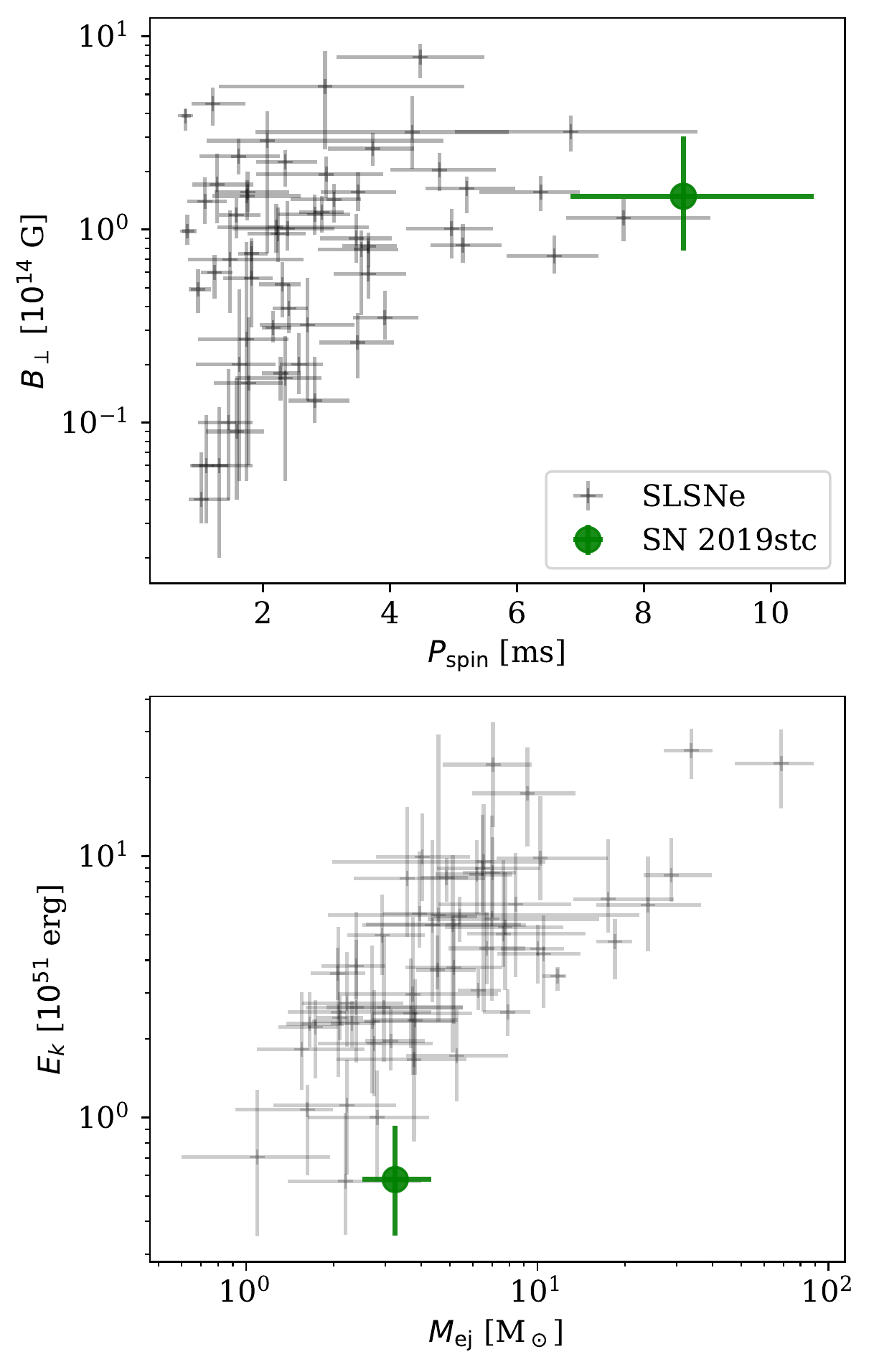}
		\caption{The best fit magnetar engine parameters for SN\,2019stc (green) compared to those of SLSNe (black; \citealt{Nicholl17,Villar18,Blanchard20}). As expected from the moderate luminosity of SN\,2019stc, the magnetar has a slower initial spin than most SLSNe, but the parameters do overlap the known distribution. \label{fig:nicholl}}
	\end{center}
\end{figure}

In Figure~\ref{fig:nicholl} we compare the magnetar engine parameters of SN\,2019stc to the SLSN samples from \citet{Nicholl17}, \citet{Villar18}, and \citet{Blanchard20}, which were also modeled with MOSFiT. We find that the spin period for SN\,2019stc is in the slow end of the distribution for SLSNe, as expected from its moderate peak luminosity compared to the bulk of the SLSN sample. Still, the magnetar engine parameters for SN\,2019stc are not outside of the distribution for SLSNe.

We therefore conclude that the first peak of SN\,2019stc is best explained with a predominant magnetar engine contribution, but with a non-zero contribution from radioactive heating. This combination explains the appearance of both the first peak of the light curve and the spectra.

\subsection{Second Light Curve Peak: Circumstellar Interaction}
\label{sec:second}

\begin{figure}[t!]
	\begin{center}
		\includegraphics[width=\columnwidth]{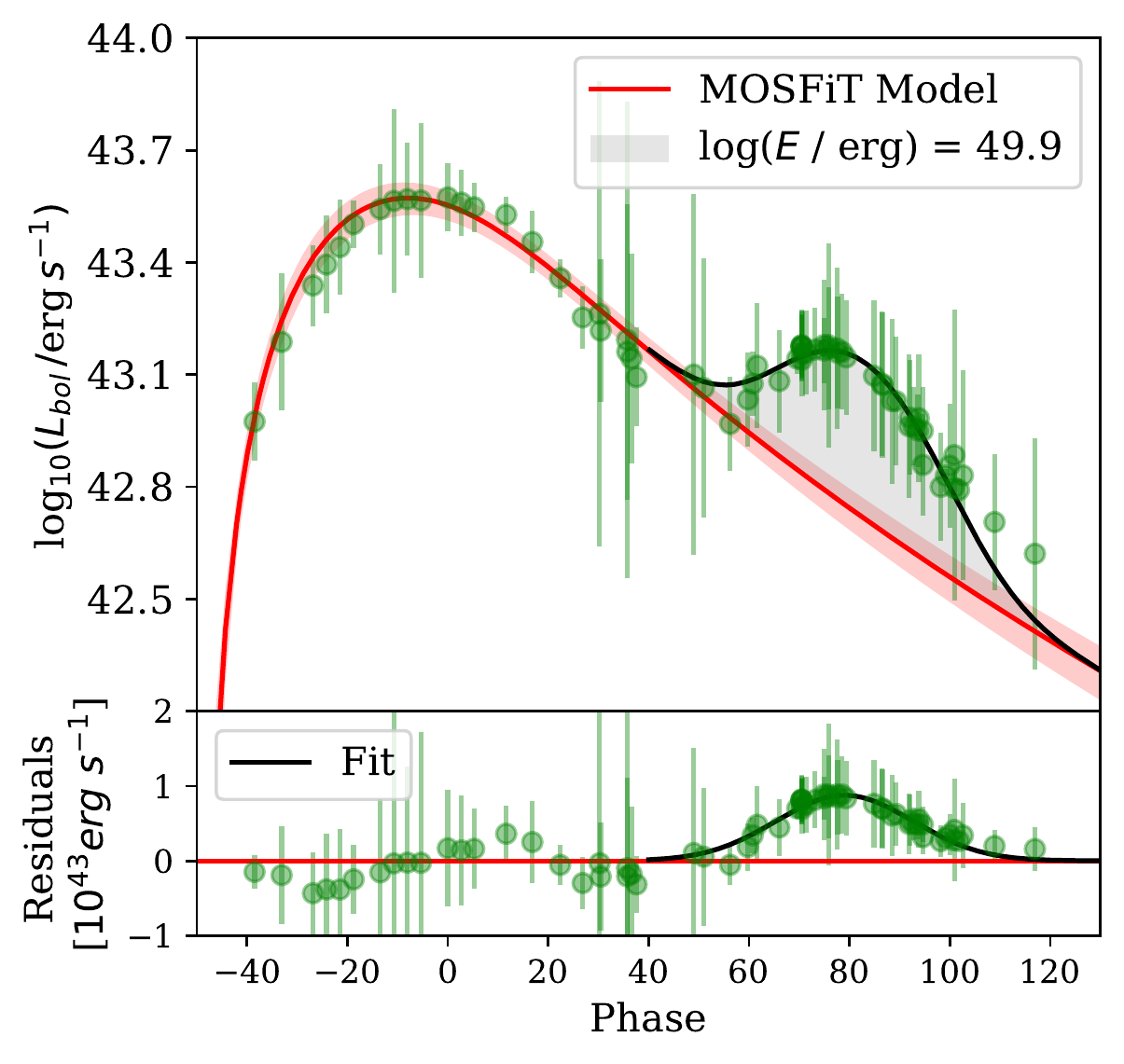}
		\caption{{\it Top:} Bolometric light curve of SN\,2019stc (green points) along with the best-fit magnetar engine plus radioactive decay model realizations (red band). The shaded grey region indicates the integrated energy in the second peak, $E_{\rm rad}\approx 8.5\times 10^{49}$ erg. {\it Bottom:} Residuals of the bolometric light curve relative to the first peak model, demonstrating that the second peak is well modeled by a Gaussian profile (black). \label{fig:modelCSM}}
	\end{center}
\end{figure}

We have so far excluded the second peak from our model fits since a single power source (or even a combined magnetar plus radioactive decay power source) cannot naturally produce a double-peaked light curve structure. To investigate the origin of the second peak we subtract the best fit magnetar plus radioactive heating model from the bolometric light curve of SN\,2019stc; see Figure \ref{fig:modelCSM}. We find that the residuals are well modeled as a Gaussian profile, with a center phase of $79.0\pm 0.4$ days, a width of ${\rm FWHM}=30.9\pm 1.2$ days, and a total integrated radiated energy of $(8.5\pm 0.3)\times 10^{49}$ erg. We stress that the energy radiated in the second peak alone is larger than for most normal SNe Ic \citep{Prentice16}.

We find that the decline rate past the second peak is $\sim 0.04$ mag day$^{-1}$, much more rapid than the decline rate for Co decay ($\sim 0.01$ mag day$^{-1}$). Therefore, the second peak is unlikely to be caused by radioactive decay. Moreover, to produce such a pronounced excess of radiated energy at late time, with a peak luminosity of $\approx 10^{43}$ erg s$^{-1}$, would require a prohibitive mass of radioactive material ($M_{\rm Ni}\approx 1.5$ M$_\odot$) buried deep in the ejecta so as not to produce detectable emission during the first peak. This is inconsistent with our finding that the photospheric radius is comparable from the first to the second peak.

We investigate the possibility that the second peak is powered by a magnetar engine with an unusually slow spin-down timescale (i.e., in this case the first peak would be dominated by radioactive heating, a model that we already disfavored in \S\ref{sec:radioactive}). However, we find that the very slow rise time of $\sim 120$ days required in this context is inconsistent with the fast decline time of $\sim 20$ days after the second peak.

\begin{figure}[t!]
	\begin{center}
		\includegraphics[width=\columnwidth]{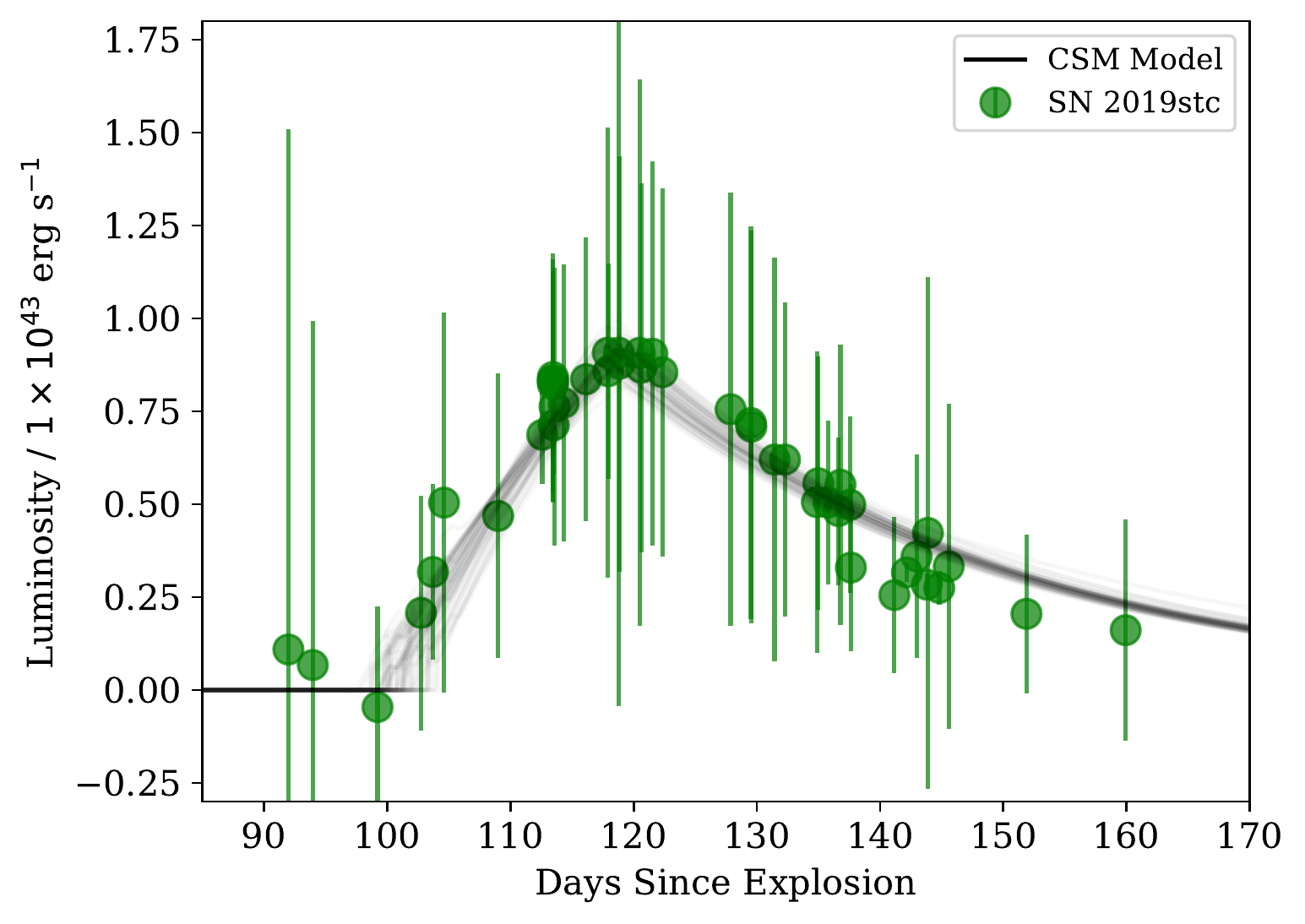}
		\caption{Residual bolometric light curve of the second peak (Figure~\ref{fig:magnetar}), along with the best-fit realizations of a CSM interaction model (grey). We find that this model can capture the overall shape of the second peak. \label{fig:residuals_magnetar}}
	\end{center}
\end{figure}

Having ruled out radioactive decay and a magnetar engine, this leaves the possibility that the second peak is powered by CSM interaction. To explore this model we fit the residual bolometric light curve with a modified version of the CSM interaction model of \citet{Chatzopoulos13} that allows the CSM to be a detached shell located at an arbitrary distance from the progenitor. The allowed density profiles of the ejecta range from $\rho\propto R$ to $\rho\propto R^{-2}$ and are described in \cite{Jiang20}. We use the relevant parameters of the best-fit magnetar plus radioactive decay model (\S\ref{sec:magnetar}) as fixed inputs: $V_{\rm ej} = 6970$ km s$^{-1}$ and $M_{\rm ej}=4$ M$_\odot$. We further assume that the CSM inside the detached shell has a wind-like density profile of $\rho\propto R^{-2}$. The remaining free parameters of the model are the CSM mass ($M_{\rm CSM}$), the opacity ($\kappa$), the inner radius of the CSM shell ($R_0$), and the density at $R_0$ ($\rho_0$). In Figure~\ref{fig:residuals_magnetar} we show the best fit realizations of this model, which give $M_{\rm CSM}=0.7\pm 0.2$ M$_\odot$, $\kappa = 0.23\pm 0.05$, $R_0 = 400\pm 11$ AU, and $\log(\rho_0 / {\rm g\ } {\rm cm}^{-3})=-13.97\pm 0.05$.

We stress that given the normal SN Ic spectrum during the second peak, the CSM shell must be hydrogen and helium poor. With a radius of about $400$ AU, and assuming a CSM velocity of $\sim 10^3$ km s$^{-1}$, expected for a compact stripped star, we find that the shell must have been ejected about 2 years before the SN explosion. If powered by CSM interaction, the lack of narrow emission features during the second peak might be the product of either the composition of the ejecta or an aspherical CSM. \citep{Soumagnac20} showed that an increasing photospheric radius is observed in SNe with aspherical CSMs, and that asphericity is more common in brighter SNe; both features we observe in SN\,2019stc.

Even though we disfavor the radioactive heating model, for completeness we perform the same CSM modeling using the parameters from this model. The relevant values from the radioactive decay model presented in \S\ref{sec:radioactive} are: $V_{\rm ej}=8000$ km s$^{-1}$, $M_{\rm ej}=10.3$ M$_\odot$, and $\kappa = 0.07$. For this model we find a larger CSM mass, $M_{\rm CSM}=3.4 \pm 0.1$ M$_\odot$, a larger inner radius, $R_0 = 490 \pm 5$ AU, and a lower density of $\log(\rho_0 / {\rm g\ } {\rm cm}^{-3})=-13.34 \pm 0.02$.

\section{Host Galaxy} 
\label{sec:host} 

Archival PS1/$3\pi$ images reveal a galaxy near the location of SN\,2019stc. Using our deep late-time images (which contain no detectable SN light contribution), we measure a half-light radius of $0.68\pm 0.15''$ for this galaxy and a projected spatial offset of $0.74\pm 0.05''$ (Figure~\ref{fig:image}). We follow the method described in \citet{Bloom02} and \citet{Berger10} to determine that this galaxy has a probability of chance coincidence of $P_{\rm cc} = 0.004$. In conjunction with the fact that the SN features agree with the host redshift, this means that this galaxy is clearly associated with SN\,2019stc. We measure a redshift of $z=0.117\pm 0.001$ by measuring the position of the narrow host galaxy emission lines H$\alpha$, H$\beta$, \mbox{[\ion{O}{2}] $\lambda\lambda$3726,3729}, and \mbox{[\ion{O}{3}] $\lambda\lambda$4959,5007}. The relevant host galaxy parameters are listed in Table~\ref{tab:host}.

We measure a flux ratio of \mbox{$L_{\rm H\alpha}/L_{\rm H\beta}=2.3\pm 0.8$} through a weighted average of all SN spectra. This value is consistent with case B recombination ($L_{\rm H\alpha}/L_{\rm H\beta} = 2.86$) indicating no significant host galaxy extinction. This agrees with the results of our MOSFiT modeling (Tables~\ref{tab:radioactive} and \ref{tab:magnetar}). From the flux of the H$\alpha$ line we estimate the star formation rate to be ${\rm SFR} = (0.11 \pm 0.01)$ M$_\odot$ yr$^{-1}$ using the relation of \citet{Kennicutt98}. 

We infer the metallicity using the $R_{23}$ method \citep{kobulnicky99}, which is double-valued, but the lack of detectable [\ion{N}{2}] emission points to the lower branch solution. This gives a value of \mbox{$12 + \log($O/H$) = 8.1 \pm 0.1$}, or \mbox{$Z = 0.26\pm 0.07$ Z$_\odot$} \citep{asplund09}. This is an unusually low metallicity for SN Ic hosts, higher than only 7\% ($N=2$) of the normal SNe Ic from the study of \citet{Modjaz20} when comparing to their KD02 calibration \citep{Kewley02}. Instead, the low metallicity is more typical of SLSN hosts, which cluster around \mbox{$12 + \log($O/H$) = 8.4\pm 0.3$} \citep{Lunnan14}. Further supporting the magnetar engine interpretation for SN\,2019stc.

\section{Discussion and Conclusions} 
\label{sec:conclusion}

\begin{deluxetable}{ccc}
    \tablecaption{Host Galaxy Properties \label{tab:host}}
    \tablehead{\colhead{}           & \colhead{Value}           & \colhead{Units}}
    \startdata
    $P_{\rm cc}$                & 0.004               &                        \\
    $z$                         & $0.117 \pm 0.001$   &                        \\
    $D_L$                       & $562 \pm 5$         & Mpc                    \\
    $D$                         & $0.74\pm 0.05$      & arcsec                 \\
    $D$                         & $2.0\pm 0.1$        & kpc                    \\
    $r_{50}$                    & $1.85 \pm 0.55$     & kpc                    \\
    $g$                         & $ 22.29 \pm 0.09$   & mag                    \\
    $r$                         & $ 22.09 \pm 0.05$   & mag                    \\
    $i$                         & $ 21.99 \pm 0.05$   & mag                    \\
    $z$                         & $ 22.04 \pm 0.10$   & mag                    \\
    $y$                         & $ 21.90 \pm 0.21$   & mag                    \\
    $M_g$                       & $ -16.67 \pm 0.09$  & mag                    \\
    $M_r$                       & $ -16.77 \pm 0.05$  & mag                    \\
    $M_i$                       & $ -16.82 \pm 0.05$  & mag                    \\
    $M_z$                       & $ -16.72 \pm 0.10$  & mag                    \\
    $M_y$                       & $ -16.84 \pm 0.21$  & mag                    \\
    $L_g$                       & $0.0348 \pm 0.0030$ & $L_*$                  \\
    $L_r$                       & $0.0128 \pm 0.0007$ & $L_*$                  \\
    $L_i$                       & $0.0109 \pm 0.0006$ & $L_*$                  \\
    $L_z$                       & $0.0064 \pm 0.0006$ & $L_*$                  \\
    $L_{\rm H\alpha}$           & $14.3 \pm 1.3$      & $10^{39}$ erg s$^{-1}$ \\
    $L_{\rm H\beta}$            & $ 6.1 \pm 1.1$      & $10^{39}$ erg s$^{-1}$ \\
    $L_{\rm [OII]_{3726,3729}}$ & $13.3 \pm 3.3$      & $10^{39}$ erg s$^{-1}$ \\
    $L_{\rm [OIII]_{4959}}$     & $ 7.6 \pm 1.8$      & $10^{39}$ erg s$^{-1}$ \\
    $L_{\rm [OIII]_{5007}}$     & $21.6 \pm 5.3$      & $10^{39}$ erg s$^{-1}$ \\
    $12 + \log($O/H$)$          & $8.1 \pm 0.1$       &                        \\
    $Z_{23}$                    & $0.26 \pm 0.07$     & Z$_\odot$              \\
    SFR                         & $0.11 \pm 0.01$     & M$_\odot$ yr$^{-1}$    \\
        \enddata
    \tablecomments{List of measured and derived properties of the host galaxy of SN\,2019stc. $P_{\rm cc}$ is the probability of chance coincidence, $z$ is the redshift, $D_L$ is the luminosity distance, $D$ is the offset of SN\,2019stc from the center of its host galaxy, $r_{50}$ is the half-light Petrosian radius in the $r$-band, $grizy$ are host magnitudes before correcting for Galactic extinction, $M_{grizy}$ are the host absolute magnitudes after correcting for Galactic extinction and including a cosmological K-correction, $L_{griz}$ are the host luminosities relative to $L_*$ in each band \citep{montero09}, $L_{n}$ are the integrated luminosities of several key emission lines (H$\alpha$, H$\beta$, [\ion{O}{2}], [\ion{O}{3}]), $12 + \log($O/H$)$, $Z_{23}$ are values for the host metallicity inferred from the R23 formalism, and SFR is the star formation rate.}
\end{deluxetable}

We presented optical photometry and spectroscopy of SN\,2019stc, which we determine to be a Type Ic SN at $z = 0.117$ with a luminous, broad, and peculiar double-peaked light curve. The two light curve peaks are separated by about 80 days and are both more prominent and further apart than in other stripped SNe with distinct peaks. From analytic modeling we conclude that SN\,2019stc is powered by three distinct power sources: a magnetar central engine, radioactive decay, and CSM interaction, with the first two powering the first peak, and the latter producing the second peak. Another example of a SN suggested to be powered by radioactive decay, a magnetar central engine, and CSM interaction was iPTF13ehe \citep{Wang16}.

Compared to the overall population of SLSNe, the magnetar engine in SN\,2019stc is much less powerful ($P_{\rm spin}\sim 7$ ms and $B_{\perp}\sim 10^{14}$ G), and this is expected to lead to a lower temperature that would not generate the distinctively blue spectrum of SLSNe near peak. This, along with a relatively higher contribution from radioactive heating compared to the more luminous SLSNe, can explain the SN Ic spectrum of SN\,2019stc near peak. Indeed, the spectrum of SN\,2019stc near peak resembles SLSN spectra well after peak, once the engine power has declined and the ejecta have significantly cooled (e.g., SN\,2015bn: \citealt{Nicholl19_nebular}, SN\,2010gx: \citealt{Pastorello10}). Thus, the spectral appearance and first peak light curve characteristics of SN\,2019stc appear to be consistent within our magnetar engine plus radioactive heating model.

We find that the prominent second peak of SN\,2019stc is instead powered by interaction with $\sim 0.7$ M$_\odot$ of hydrogen-free CSM located about 400 AU from the progenitor (corresponding to ejection about 2 years pre-explosion). Although only a handful of SLSNe show possible evidence for late-time hydrogen emission \citep{Yan20}, the light curve ``bumps'' in other SLSNe have been suggested to result from CSM interaction (e.g., \citealt{Nicholl16_15bn,Inserra17,Blanchard18,Lunnan20,Jin21}). These bumps, however, are significantly less pronounced than in SN\,2019stc, and the inferred CSM masses required to produce them are $\lesssim 0.05$ M$_\odot$, more than an order of magnitude lower than in SN\,2019stc.

Combining the ejecta mass of $\approx 4$ M$_\odot$, CSM shell mass of $\approx 0.7$ M$_\odot$, and remnant neutron star mass of $\approx 1.7$ M$_\odot$, we infer a total pre-explosion progenitor CO core mass of $\approx 6.5$ M$_\odot$. The ejecta mass for SN\,2019stc is within the range of typical ejecta masses of SNe Ic, which center around $M_{\rm ej} = 4.5\pm 0.8$ M$_\odot$ \citep{Barbarino20}, but the CO core mass also overlaps with the progenitor mass distribution for SLSNe, which have mean masses of $6.4_{-2.8}^{+8.1}$ M$_\odot$ \citep{Blanchard20}.

The large CSM density and proximity to the progenitor indicate the $0.7$ M$_\odot$ were ejected within a span of $\sim 0.1$ years about 2 years before explosion. SNe with high mass-loss events have been observed previously (e.g., SN\,2014C \citealt{Milisavljevic15};  SN\,2009ip \citealt{Margutti14}), but the CSM in these events was hydrogen-rich. Large mass-loss events are also a key feature of pulsational pair-instability SNe (PPISNe; \citealt{Woosley07}). We note however that the progenitor mass estimate of SN\,2019stc is well below the required CO core mass of $\gtrsim 25$ M$_\odot$ required to produce a PPISNe \citep{Woosley07}. 

In conclusion, SN\,2019stc is a peculiar stripped SN that occupies the mostly unexplored regime between SLSNe and SNe Ic, and shares properties with both. Its peak luminosity is on the low end for SLSNe, but high for SNe Ic and Ic-BL. Its first-peak light curve is unusual for SNe Ic but can be accommodated with the same models that explain SLSNe. Its host environment has a low metallicity, much like those of SLSNe and its pre-explosion progenitor mass is consistent with those of both SNe Ic and SLSNe.

\acknowledgments
We thank Y.~Beletsky for carrying out some of the Magellan observations. The Berger Time-Domain Group at Harvard is supported by NSF and NASA grants. This work is supported in part by the National Science Foundation under Cooperative Agreement PHY-2019786 (The NSF AI Institute for Artificial Intelligence and Fundamental Interactions, http://iaifi.org/). S.~Gomez is partly supported by an NSF Graduate Research Fellowship. This paper includes data gathered with the 6.5 meter Magellan Telescopes located at Las Campanas Observatory, Chile. Observations reported here were obtained at the MMT Observatory, a joint facility of the University of Arizona and the Smithsonian Institution. This research has made use of NASA’s Astrophysics Data System. This research has made use of the SIMBAD database, operated at CDS, Strasbourg, France. 

\software{Astropy\citep{astropy18}, MOSFiT\citep{guillochon18}, FLEET\citep{FLEET}, PyRAF\citep{science12}, SAOImage DS9\citep{Smithsonian00}, emcee\citep{foreman13}, corner\citep{foreman16}, HOTPANTS\citep{Becker15}, Superbol\citep{Nicholl18_superbol}, Matplotlib\citep{hunter07}, SciPy\citep{Walt11}, NumPy\citep{Oliphant07}, extinction\citep{Barbary16}, PYPHOT(\url{https://github.com/mfouesneau/pyphot}}).

\bibliography{references}

\begin{thebibliography}{}
\expandafter\ifx\csname natexlab\endcsname\relax\def\natexlab#1{#1}\fi
\providecommand{\url}[1]{\href{#1}{#1}}
\providecommand{\dodoi}[1]{doi:~\href{http://doi.org/#1}{\nolinkurl{#1}}}

\bibitem[{{Anderson} {et~al.}(2018){Anderson}, {Pessi}, {Dessart}, {Inserra},
  {Hiramatsu}, {Taggart}, {Smartt}, {Leloudas}, {Chen}, {M{\"o}ller}, {Roy},
  {Schulze}, {Perley}, {Selsing}, {Prentice}, {Gal-Yam}, {Angus}, {Arcavi},
  {Ashall}, {Bulla}, {Bray}, {Burke}, {Callis}, {Cartier}, {Chang}, {Chambers},
  {Clark}, {Denneau}, {Dennefeld}, {Flewelling}, {Fraser}, {Galbany},
  {Gromadzki}, {Guti{\'e}rrez}, {Heinze}, {Hosseinzadeh}, {Howell}, {Hsiao},
  {Kankare}, {Kostrzewa-Rutkowska}, {Magnier}, {Maguire}, {Mazzali}, {McBrien},
  {McCully}, {Morrell}, {Lowe}, {Onken}, {Onori}, {Phillips}, {Rest},
  {Ridden-Harper}, {Ruiter}, {Sand}, {Smith}, {Smith}, {Stalder},
  {Stritzinger}, {Sullivan}, {Tonry}, {Tucker}, {Valenti}, {Wainscoat},
  {Waters}, {Wolf}, \& {Young}}]{Anderson18}
{Anderson}, J.~P., {Pessi}, P.~J., {Dessart}, L., {et~al.} 2018,
  \hypersetup{urlcolor=magenta}\href{https://dx.doi.org/10.1051/0004-6361/201833725}{\aap},
  \hypersetup{urlcolor=blue}\href{https://ui.adsabs.harvard.edu/abs/2018A&A...620A..67A}{620,
  A67}

\bibitem[{{Angus} {et~al.}(2019){Angus}, {Smith}, {Sullivan}, {Inserra},
  {Wiseman}, {D'Andrea}, {Thomas}, {Nichol}, {Galbany}, {Childress}, {Asorey},
  {Brown}, {Casas}, {Castander}, {Curtin}, {Frohmaier}, {Glazebrook}, {Gruen},
  {Gutierrez}, {Kessler}, {Kim}, {Lidman}, {Macaulay}, {Nugent}, {Pursiainen},
  {Sako}, {Soares-Santos}, {Thomas}, {Abbott}, {Avila}, {Bertin}, {Brooks},
  {Buckley-Geer}, {Burke}, {Carnero Rosell}, {Carretero}, {da Costa}, {De
  Vicente}, {Desai}, {Diehl}, {Doel}, {Eifler}, {Flaugher}, {Fosalba},
  {Frieman}, {Garc{\'\i}a-Bellido}, {Gruendl}, {Gschwend}, {Hartley},
  {Hollowood}, {Honscheid}, {Hoyle}, {James}, {Kuehn}, {Kuropatkin}, {Lahav},
  {Lima}, {Maia}, {March}, {Marshall}, {Menanteau}, {Miller}, {Miquel},
  {Ogando}, {Plazas}, {Romer}, {Sanchez}, {Schindler}, {Schubnell}, {Sobreira},
  {Suchyta}, {Swanson}, {Tarle}, {Thomas}, {Tucker}, \& {DES
  Collaboration}}]{Angus19}
{Angus}, C.~R., {Smith}, M., {Sullivan}, M., {et~al.} 2019,
  \hypersetup{urlcolor=magenta}\href{https://dx.doi.org/10.1093/mnras/stz1321}{\mnras},
  \hypersetup{urlcolor=blue}\href{https://ui.adsabs.harvard.edu/abs/2019MNRAS.487.2215A}{487,
  2215}

\bibitem[{{Arnett}(1982)}]{Arnett82}
{Arnett}, W.~D. 1982,
  \hypersetup{urlcolor=magenta}\href{https://dx.doi.org/10.1086/159681}{\apj},
  \hypersetup{urlcolor=blue}\href{https://ui.adsabs.harvard.edu/abs/1982ApJ...253..785A}{253,
  785}

\bibitem[{{Asplund} {et~al.}(2009){Asplund}, {Grevesse}, {Sauval}, \&
  {Scott}}]{asplund09}
{Asplund}, M., {Grevesse}, N., {Sauval}, A.~J., \& {Scott}, P. 2009,
  \hypersetup{urlcolor=magenta}\href{https://dx.doi.org/10.1146/annurev.astro.46.060407.145222}{\araa},
  \hypersetup{urlcolor=blue}\href{https://ui.adsabs.harvard.edu/abs/2009ARA&A..47..481A}{47,
  481}

\bibitem[{{Astropy Collaboration} {et~al.}(2018){Astropy Collaboration},
  {Price-Whelan}, {Sip{\H{o}}cz}, {G{\"u}nther}, {Lim}, {Crawford}, {Conseil},
  {Shupe}, {Craig}, \& {Dencheva}}]{astropy18}
{Astropy Collaboration}, {Price-Whelan}, A.~M., {Sip{\H{o}}cz}, B.~M., {et~al.}
  2018,
  \hypersetup{urlcolor=magenta}\href{https://dx.doi.org/10.3847/1538-3881/aabc4f}{\aj},
  \hypersetup{urlcolor=blue}\href{https://ui.adsabs.harvard.edu/abs/2018AJ....156..123A}{156,
  123}

\bibitem[{{Barbarino} {et~al.}(2020){Barbarino}, {Sollerman}, {Taddia},
  {Fremling}, {Karamehmetoglu}, {Arcavi}, {Gal-Yam}, {Laher}, {Schulze},
  {Wozniak}, \& {Yan}}]{Barbarino20}
{Barbarino}, C., {Sollerman}, J., {Taddia}, F., {et~al.} 2020, arXiv e-prints,
  \hypersetup{urlcolor=magenta}\href{https://arxiv.org/abs/2010.08392}{arXiv}{:}\hypersetup{urlcolor=blue}\href{https://ui.adsabs.harvard.edu/abs/2020arXiv201008392B}{2010.08392}

\bibitem[{Barbary(2016)}]{Barbary16}
Barbary, K. 2016, extinction, v0.3.0,  Zenodo,
  \hypersetup{urlcolor=magenta}doi:\href{https://dx.doi.org/10.5281/zenodo.804967}{10.5281/zenodo.804967}

\bibitem[{{Becker}(2015)}]{Becker15}
{Becker}, A. 2015, {HOTPANTS: High Order Transform of PSF ANd Template
  Subtraction},
  \hypersetup{urlcolor=magenta}\href{https://ascl.net/1504.004}{ascl}:\hypersetup{urlcolor=blue}\href{https://ui.adsabs.harvard.edu/abs/2015ascl.soft04004B}{1504.004}

\bibitem[{{Bellm} {et~al.}(2019){Bellm}, {Kulkarni}, {Graham}, {Dekany},
  {Smith}, {Riddle}, {Masci}, {Helou}, {Prince}, {Adams}, {Barbarino},
  {Barlow}, {Bauer}, {Beck}, {Belicki}, {Biswas}, {Blagorodnova}, {Bodewits},
  {Bolin}, {Brinnel}, {Brooke}, {Bue}, {Bulla}, {Burruss}, {Cenko}, {Chang},
  {Connolly}, {Coughlin}, {Cromer}, {Cunningham}, {De}, {Delacroix}, {Desai},
  {Duev}, {Eadie}, {Farnham}, {Feeney}, {Feindt}, {Flynn}, {Franckowiak},
  {Frederick}, {Fremling}, {Gal-Yam}, {Gezari}, {Giomi}, {Goldstein},
  {Golkhou}, {Goobar}, {Groom}, {Hacopians}, {Hale}, {Henning}, {Ho}, {Hover},
  {Howell}, {Hung}, {Huppenkothen}, {Imel}, {Ip}, {Ivezi{\'c}}, {Jackson},
  {Jones}, {Juric}, {Kasliwal}, {Kaspi}, {Kaye}, {Kelley}, {Kowalski},
  {Kramer}, {Kupfer}, {Landry}, {Laher}, {Lee}, {Lin}, {Lin}, {Lunnan},
  {Giomi}, {Mahabal}, {Mao}, {Miller}, {Monkewitz}, {Murphy}, {Ngeow},
  {Nordin}, {Nugent}, {Ofek}, {Patterson}, {Penprase}, {Porter}, {Rauch},
  {Rebbapragada}, {Reiley}, {Rigault}, {Rodriguez}, {van Roestel}, {Rusholme},
  {van Santen}, {Schulze}, {Shupe}, {Singer}, {Soumagnac}, {Stein}, {Surace},
  {Sollerman}, {Szkody}, {Taddia}, {Terek}, {Van Sistine}, {van Velzen},
  {Vestrand}, {Walters}, {Ward}, {Ye}, {Yu}, {Yan}, \& {Zolkower}}]{Bellm19}
{Bellm}, E.~C., {Kulkarni}, S.~R., {Graham}, M.~J., {et~al.} 2019,
  \hypersetup{urlcolor=magenta}\href{https://dx.doi.org/10.1088/1538-3873/aaecbe}{\pasp},
  \hypersetup{urlcolor=blue}\href{https://ui.adsabs.harvard.edu/abs/2019PASP..131a8002B}{131,
  018002}

\bibitem[{{Berger}(2010)}]{Berger10}
{Berger}, E. 2010,
  \hypersetup{urlcolor=magenta}\href{https://dx.doi.org/10.1088/0004-637X/722/2/1946}{\apj},
  \hypersetup{urlcolor=blue}\href{https://ui.adsabs.harvard.edu/abs/2010ApJ...722.1946B}{722,
  1946}

\bibitem[{{Blanchard} {et~al.}(2020){Blanchard}, {Berger}, {Nicholl}, \&
  {Villar}}]{Blanchard20}
{Blanchard}, P.~K., {Berger}, E., {Nicholl}, M., \& {Villar}, V.~A. 2020,
  \hypersetup{urlcolor=magenta}\href{https://dx.doi.org/10.3847/1538-4357/ab9638}{\apj},
  \hypersetup{urlcolor=blue}\href{https://ui.adsabs.harvard.edu/abs/2020ApJ...897..114B}{897,
  114}

\bibitem[{{Blanchard} {et~al.}(2019){Blanchard}, {Nicholl}, {Berger},
  {Chornock}, {Milisavljevic}, {Margutti}, \& {Gomez}}]{Blanchard19}
{Blanchard}, P.~K., {Nicholl}, M., {Berger}, E., {et~al.} 2019,
  \hypersetup{urlcolor=magenta}\href{https://dx.doi.org/10.3847/1538-4357/aafa13}{\apj},
  \hypersetup{urlcolor=blue}\href{https://ui.adsabs.harvard.edu/abs/2019ApJ...872...90B}{872,
  90}

\bibitem[{{Blanchard} {et~al.}(2018){Blanchard}, {Nicholl}, {Berger},
  {Chornock}, {Margutti}, {Milisavljevic}, {Fong}, {MacLeod}, \&
  {Bhirombhakdi}}]{Blanchard18}
{Blanchard}, P.~K., {Nicholl}, M., {Berger}, E., {et~al.} 2018,
  \hypersetup{urlcolor=magenta}\href{https://dx.doi.org/10.3847/1538-4357/aad8b9}{\apj},
  \hypersetup{urlcolor=blue}\href{https://ui.adsabs.harvard.edu/abs/2018ApJ...865....9B}{865,
  9}

\bibitem[{{Bloom} {et~al.}(2002){Bloom}, {Kulkarni}, \& {Djorgovski}}]{Bloom02}
{Bloom}, J.~S., {Kulkarni}, S.~R., \& {Djorgovski}, S.~G. 2002,
  \hypersetup{urlcolor=magenta}\href{https://dx.doi.org/10.1086/338893}{\aj},
  \hypersetup{urlcolor=blue}\href{https://ui.adsabs.harvard.edu/abs/2002AJ....123.1111B}{123,
  1111}

\bibitem[{{Cardelli} {et~al.}(1989){Cardelli}, {Clayton}, \&
  {Mathis}}]{Cardelli89}
{Cardelli}, J.~A., {Clayton}, G.~C., \& {Mathis}, J.~S. 1989, in Interstellar
  Dust, ed. L.~J. {Allamandola} \& A.~G.~G.~M. {Tielens}, Vol. 135, 5--10

\bibitem[{{Chambers} \& {Pan-STARRS Team}(2018)}]{Chambers18}
{Chambers}, K., \& {Pan-STARRS Team}. 2018, in American Astronomical Society
  Meeting Abstracts, Vol. 231, American Astronomical Society Meeting Abstracts
  \#231, 102.01

\bibitem[{{Chatzopoulos} {et~al.}(2013){Chatzopoulos}, {Wheeler}, {Vinko},
  {Horvath}, \& {Nagy}}]{Chatzopoulos13}
{Chatzopoulos}, E., {Wheeler}, J.~C., {Vinko}, J., {Horvath}, Z.~L., \& {Nagy},
  A. 2013,
  \hypersetup{urlcolor=magenta}\href{https://dx.doi.org/10.1088/0004-637X/773/1/76}{\apj},
  \hypersetup{urlcolor=blue}\href{https://ui.adsabs.harvard.edu/abs/2013ApJ...773...76C}{773,
  76}

\bibitem[{{Chomiuk} {et~al.}(2011){Chomiuk}, {Chornock}, {Soderberg}, {Berger},
  {Chevalier}, {Foley}, {Huber}, {Narayan}, {Rest}, \& {Gezari}}]{Chomiuk11}
{Chomiuk}, L., {Chornock}, R., {Soderberg}, A.~M., {et~al.} 2011,
  \hypersetup{urlcolor=magenta}\href{https://dx.doi.org/10.1088/0004-637X/743/2/114}{\apj},
  \hypersetup{urlcolor=blue}\href{https://ui.adsabs.harvard.edu/abs/2011ApJ...743..114C}{743,
  114}

\bibitem[{{Dressler} {et~al.}(2011){Dressler}, {Bigelow}, {Hare}, {Sutin},
  {Thompson}, {Burley}, {Epps}, {Oemler}, {Bagish}, \& {Birk}}]{dressler11}
{Dressler}, A., {Bigelow}, B., {Hare}, T., {et~al.} 2011,
  \hypersetup{urlcolor=magenta}\href{https://dx.doi.org/10.1086/658908}{\pasp},
  \hypersetup{urlcolor=blue}\href{https://ui.adsabs.harvard.edu/abs/2011PASP..123..288D}{123,
  288}

\bibitem[{{Drout} {et~al.}(2011){Drout}, {Soderberg}, {Gal-Yam}, {Cenko},
  {Fox}, {Leonard}, {Sand}, {Moon}, {Arcavi}, \& {Green}}]{Drout11}
{Drout}, M.~R., {Soderberg}, A.~M., {Gal-Yam}, A., {et~al.} 2011,
  \hypersetup{urlcolor=magenta}\href{https://dx.doi.org/10.1088/0004-637X/741/2/97}{\apj},
  \hypersetup{urlcolor=blue}\href{https://ui.adsabs.harvard.edu/abs/2011ApJ...741...97D}{741,
  97}

\bibitem[{{Fabricant} {et~al.}(2019){Fabricant}, {Fata}, {Epps}, {Gauron},
  {Mueller}, {Zajac}, {Amato}, {Barberis}, {Bergner}, {Brennan}, {Brown},
  {Chilingarian}, {Geary}, {Kradinov}, {McLeod}, {Smith}, \&
  {Woods}}]{Fabricant19}
{Fabricant}, D., {Fata}, R., {Epps}, H., {et~al.} 2019,
  \hypersetup{urlcolor=magenta}\href{https://dx.doi.org/10.1088/1538-3873/ab1d78}{\pasp},
  \hypersetup{urlcolor=blue}\href{https://ui.adsabs.harvard.edu/abs/2019PASP..131g5004F}{131,
  075004}

\bibitem[{{Fabricant} {et~al.}(2003){Fabricant}, {Epps}, {Brown}, {Fata}, \&
  {Mueller}}]{fabricant03}
{Fabricant}, D.~G., {Epps}, H.~W., {Brown}, W.~L., {Fata}, R.~G., \& {Mueller},
  M. 2003, in Society of Photo-Optical Instrumentation Engineers (SPIE)
  Conference Series, Vol. 4841, Instrument Design and Performance for
  Optical/Infrared Ground-based Telescopes, ed. M.~{Iye} \& A.~F.~M.
  {Moorwood}, 1134--1144

\bibitem[{{Filippenko}(1997)}]{Filippenko97}
{Filippenko}, A.~V. 1997,
  \hypersetup{urlcolor=magenta}\href{https://dx.doi.org/10.1146/annurev.astro.35.1.309}{\araa},
  \hypersetup{urlcolor=blue}\href{https://ui.adsabs.harvard.edu/abs/1997ARA&A..35..309F}{35,
  309}

\bibitem[{{Foreman-Mackey}(2016)}]{foreman16}
{Foreman-Mackey}, D. 2016,
  \hypersetup{urlcolor=magenta}\href{https://dx.doi.org/10.21105/joss.00024}{The
  Journal of Open Source Software},
  \hypersetup{urlcolor=blue}\href{https://ui.adsabs.harvard.edu/abs/2016JOSS....1...24F}{1}

\bibitem[{{Foreman-Mackey} {et~al.}(2013){Foreman-Mackey}, {Hogg}, {Lang}, \&
  {Goodman}}]{foreman13}
{Foreman-Mackey}, D., {Hogg}, D.~W., {Lang}, D., \& {Goodman}, J. 2013,
  \hypersetup{urlcolor=magenta}\href{https://dx.doi.org/10.1086/670067}{\pasp},
  \hypersetup{urlcolor=blue}\href{https://ui.adsabs.harvard.edu/abs/2013PASP..125..306F}{125,
  306}

\bibitem[{{Fremling} {et~al.}(2018){Fremling}, {Sollerman}, {Kasliwal},
  {Kulkarni}, {Barbarino}, {Ergon}, {Karamehmetoglu}, {Taddia}, {Arcavi},
  {Cenko}, {Clubb}, {De Cia}, {Duggan}, {Filippenko}, {Gal-Yam}, {Graham},
  {Horesh}, {Hosseinzadeh}, {Howell}, {Kuesters}, {Lunnan}, {Matheson},
  {Nugent}, {Perley}, {Quimby}, \& {Saunders}}]{Fremling18}
{Fremling}, C., {Sollerman}, J., {Kasliwal}, M.~M., {et~al.} 2018,
  \hypersetup{urlcolor=magenta}\href{https://dx.doi.org/10.1051/0004-6361/201731701}{\aap},
  \hypersetup{urlcolor=blue}\href{https://ui.adsabs.harvard.edu/abs/2018A&A...618A..37F}{618,
  A37}

\bibitem[{{Gal-Yam}(2012)}]{Gal-Yam12}
{Gal-Yam}, A. 2012,
  \hypersetup{urlcolor=magenta}\href{https://dx.doi.org/10.1126/science.1203601}{Science},
  \hypersetup{urlcolor=blue}\href{https://ui.adsabs.harvard.edu/abs/2012Sci...337..927G}{337,
  927}

\bibitem[{{Gal-Yam}(2019)}]{Gal-Yam19}
{Gal-Yam}, A. 2019,
  \hypersetup{urlcolor=magenta}\href{https://dx.doi.org/10.1146/annurev-astro-081817-051819}{\araa},
  \hypersetup{urlcolor=blue}\href{https://ui.adsabs.harvard.edu/abs/2019ARA&A..57..305G}{57,
  305}

\bibitem[{{Galama} {et~al.}(1998){Galama}, {Vreeswijk}, {van Paradijs},
  {Kouveliotou}, {Augusteijn}, {B{\"o}hnhardt}, {Brewer}, {Doublier},
  {Gonzalez}, \& {Leibundgut}}]{galama98}
{Galama}, T.~J., {Vreeswijk}, P.~M., {van Paradijs}, J., {et~al.} 1998,
  \hypersetup{urlcolor=magenta}\href{https://dx.doi.org/10.1038/27150}{\nat},
  \hypersetup{urlcolor=blue}\href{https://ui.adsabs.harvard.edu/abs/1998Natur.395..670G}{395,
  670}

\bibitem[{{Gelman} \& {Rubin}(1992)}]{gelman92}
{Gelman}, A., \& {Rubin}, D.~B. 1992,
  \hypersetup{urlcolor=magenta}\href{https://dx.doi.org/10.1214/ss/1177011136}{Statistical
  Science},
  \hypersetup{urlcolor=blue}\href{https://ui.adsabs.harvard.edu/abs/1992StaSc...7..457G}{7,
  457}

\bibitem[{{Gomez} {et~al.}(2020{\natexlab{\hspace{0pt}a}}){Gomez}, {Berger},
  {Blanchard}, {Hosseinzadeh}, {Nicholl}, {Villar}, \& {Yin}}]{Gomez20}
{Gomez}, S., {Berger}, E., {Blanchard}, P.~K., {et~al.}
  2020{\natexlab{\hspace{0pt}a}},
  \hypersetup{urlcolor=magenta}\href{https://dx.doi.org/10.3847/1538-4357/abbf49}{\apj},
  \hypersetup{urlcolor=blue}\href{https://ui.adsabs.harvard.edu/abs/2020ApJ...904...74G}{904,
  74}

\bibitem[{{Gomez} {et~al.}(2020{\natexlab{\hspace{0pt}b}}){Gomez}, {Berger},
  {Blanchard}, {Hosseinzadeh}, {Nicholl}, {Villar}, \& {Yin}}]{FLEET}
{Gomez}, S., {Berger}, E., {Blanchard}, P.~K., {et~al.}
  2020{\natexlab{\hspace{0pt}b}}, {FLEET Finding Luminous and Exotic
  Extragalactic Transients}, 1.0.0,  Zenodo,
  \hypersetup{urlcolor=magenta}doi:\href{https://dx.doi.org/10.5281/zenodo.4013965}{10.5281/zenodo.4013965}

\bibitem[{{Gomez} {et~al.}(2019){Gomez}, {Berger}, {Nicholl}, {Blanchard},
  {Villar}, {Patton}, {Chornock}, {Leja}, {Hosseinzadeh}, \&
  {Cowperthwaite}}]{Gomez19}
{Gomez}, S., {Berger}, E., {Nicholl}, M., {et~al.} 2019,
  \hypersetup{urlcolor=magenta}\href{https://dx.doi.org/10.3847/1538-4357/ab2f92}{\apj},
  \hypersetup{urlcolor=blue}\href{https://ui.adsabs.harvard.edu/abs/2019ApJ...881...87G}{881,
  87}

\bibitem[{{Greiner} {et~al.}(2015){Greiner}, {Mazzali}, {Kann}, {Kr{\"u}hler},
  {Pian}, {Prentice}, {Olivares E.}, {Rossi}, {Klose}, {Taubenberger}, {Knust},
  {Afonso}, {Ashall}, {Bolmer}, {Delvaux}, {Diehl}, {Elliott}, {Filgas},
  {Fynbo}, {Graham}, {Guelbenzu}, {Kobayashi}, {Leloudas}, {Savaglio},
  {Schady}, {Schmidl}, {Schweyer}, {Sudilovsky}, {Tanga}, {Updike}, {van
  Eerten}, \& {Varela}}]{Greiner15}
{Greiner}, J., {Mazzali}, P.~A., {Kann}, D.~A., {et~al.} 2015,
  \hypersetup{urlcolor=magenta}\href{https://dx.doi.org/10.1038/nature14579}{\nat},
  \hypersetup{urlcolor=blue}\href{https://ui.adsabs.harvard.edu/abs/2015Natur.523..189G}{523,
  189}

\bibitem[{{Guillochon} {et~al.}(2018){Guillochon}, {Nicholl}, {Villar},
  {Mockler}, {Narayan}, {Mandel}, {Berger}, \& {Williams}}]{guillochon18}
{Guillochon}, J., {Nicholl}, M., {Villar}, V.~A., {et~al.} 2018,
  \hypersetup{urlcolor=magenta}\href{https://dx.doi.org/10.3847/1538-4365/aab761}{\apjs},
  \hypersetup{urlcolor=blue}\href{https://ui.adsabs.harvard.edu/abs/2018ApJS..236....6G}{236,
  6}

\bibitem[{{Guillochon} {et~al.}(2017){Guillochon}, {Parrent}, {Kelley}, \&
  {Margutti}}]{guillochon17}
{Guillochon}, J., {Parrent}, J., {Kelley}, L.~Z., \& {Margutti}, R. 2017,
  \hypersetup{urlcolor=magenta}\href{https://dx.doi.org/10.3847/1538-4357/835/1/64}{\apj},
  \hypersetup{urlcolor=blue}\href{https://ui.adsabs.harvard.edu/abs/2017ApJ...835...64G}{835,
  64}

\bibitem[{{Hinshaw} {et~al.}(2013){Hinshaw}, {Larson}, {Komatsu}, {Spergel},
  {Bennett}, {Dunkley}, {Nolta}, {Halpern}, {Hill}, \& {Odegard}}]{hinshaw13}
{Hinshaw}, G., {Larson}, D., {Komatsu}, E., {et~al.} 2013,
  \hypersetup{urlcolor=magenta}\href{https://dx.doi.org/10.1088/0067-0049/208/2/19}{\apjs},
  \hypersetup{urlcolor=blue}\href{https://ui.adsabs.harvard.edu/abs/2013ApJS..208...19H}{208,
  19}

\bibitem[{{Ho} {et~al.}(2019){Ho}, {Goldstein}, {Schulze}, {Khatami}, {Perley},
  {Ergon}, {Gal-Yam}, {Corsi}, {Andreoni}, {Barbarino}, {Bellm},
  {Blagorodnova}, {Bright}, {Burns}, {Cenko}, {Cunningham}, {De}, {Dekany},
  {Dugas}, {Fender}, {Fransson}, {Fremling}, {Goldstein}, {Graham}, {Hale},
  {Horesh}, {Hung}, {Kasliwal}, {Kuin}, {Kulkarni}, {Kupfer}, {Lunnan},
  {Masci}, {Ngeow}, {Nugent}, {Ofek}, {Patterson}, {Petitpas}, {Rusholme},
  {Sai}, {Sfaradi}, {Shupe}, {Sollerman}, {Soumagnac}, {Tachibana}, {Taddia},
  {Walters}, {Wang}, {Yao}, \& {Zhang}}]{Ho19}
{Ho}, A. Y.~Q., {Goldstein}, D.~A., {Schulze}, S., {et~al.} 2019,
  \hypersetup{urlcolor=magenta}\href{https://dx.doi.org/10.3847/1538-4357/ab55ec}{\apj},
  \hypersetup{urlcolor=blue}\href{https://ui.adsabs.harvard.edu/abs/2019ApJ...887..169H}{887,
  169}

\bibitem[{{Howell} {et~al.}(2013){Howell}, {Kasen}, {Lidman}, {Sullivan},
  {Conley}, {Astier}, {Balland}, {Carlberg}, {Fouchez}, {Guy}, {Hardin},
  {Pain}, {Palanque-Delabrouille}, {Perrett}, {Pritchet}, {Regnault}, {Rich},
  \& {Ruhlmann-Kleider}}]{Howell13}
{Howell}, D.~A., {Kasen}, D., {Lidman}, C., {et~al.} 2013,
  \hypersetup{urlcolor=magenta}\href{https://dx.doi.org/10.1088/0004-637X/779/2/98}{\apj},
  \hypersetup{urlcolor=blue}\href{https://ui.adsabs.harvard.edu/abs/2013ApJ...779...98H}{779,
  98}

\bibitem[{{Hunter}(2007)}]{hunter07}
{Hunter}, J.~D. 2007,
  \hypersetup{urlcolor=magenta}\href{https://dx.doi.org/10.1109/MCSE.2007.55}{Computing
  in Science and Engineering},
  \hypersetup{urlcolor=blue}\href{https://ui.adsabs.harvard.edu/abs/2007CSE.....9...90H}{9,
  90}

\bibitem[{{Inserra} {et~al.}(2017){Inserra}, {Nicholl}, {Chen}, {Jerkstrand},
  {Smartt}, {Kr{\"u}hler}, {Anderson}, {Baltay}, {Della Valle}, {Fraser},
  {Gal-Yam}, {Galbany}, {Kankare}, {Maguire}, {Rabinowitz}, {Smith}, {Valenti},
  \& {Young}}]{Inserra17}
{Inserra}, C., {Nicholl}, M., {Chen}, T.~W., {et~al.} 2017,
  \hypersetup{urlcolor=magenta}\href{https://dx.doi.org/10.1093/mnras/stx834}{\mnras},
  \hypersetup{urlcolor=blue}\href{https://ui.adsabs.harvard.edu/abs/2017MNRAS.468.4642I}{468,
  4642}

\bibitem[{{Jiang} {et~al.}(2020){Jiang}, {Jiang}, \& {Ashley Villar}}]{Jiang20}
{Jiang}, B., {Jiang}, S., \& {Ashley Villar}, V. 2020,
  \hypersetup{urlcolor=magenta}\href{https://dx.doi.org/10.3847/2515-5172/ab7128}{Research
  Notes of the American Astronomical Society},
  \hypersetup{urlcolor=blue}\href{https://ui.adsabs.harvard.edu/abs/2020RNAAS...4...16J}{4,
  16}

\bibitem[{{Jin} {et~al.}(2021){Jin}, {Yoon}, \& {Blinnikov}}]{Jin21}
{Jin}, H., {Yoon}, S.-C., \& {Blinnikov}, S. 2021, arXiv e-prints,
  \hypersetup{urlcolor=magenta}\href{https://arxiv.org/abs/2101.11171}{arXiv}{:}\hypersetup{urlcolor=blue}\href{https://ui.adsabs.harvard.edu/abs/2021arXiv210111171J}{2101.11171}

\bibitem[{{Kasen} \& {Bildsten}(2010)}]{Kasen10}
{Kasen}, D., \& {Bildsten}, L. 2010,
  \hypersetup{urlcolor=magenta}\href{https://dx.doi.org/10.1088/0004-637X/717/1/245}{\apj},
  \hypersetup{urlcolor=blue}\href{https://ui.adsabs.harvard.edu/abs/2010ApJ...717..245K}{717,
  245}

\bibitem[{{Kennicutt}(1998)}]{Kennicutt98}
{Kennicutt}, Robert~C., J. 1998,
  \hypersetup{urlcolor=magenta}\href{https://dx.doi.org/10.1146/annurev.astro.36.1.189}{\araa},
  \hypersetup{urlcolor=blue}\href{https://ui.adsabs.harvard.edu/abs/1998ARA&A..36..189K}{36,
  189}

\bibitem[{{Kewley} \& {Dopita}(2002)}]{Kewley02}
{Kewley}, L.~J., \& {Dopita}, M.~A. 2002,
  \hypersetup{urlcolor=magenta}\href{https://dx.doi.org/10.1086/341326}{\apjs},
  \hypersetup{urlcolor=blue}\href{https://ui.adsabs.harvard.edu/abs/2002ApJS..142...35K}{142,
  35}

\bibitem[{{Kobulnicky} {et~al.}(1999){Kobulnicky}, {Kennicutt}, \&
  {Pizagno}}]{kobulnicky99}
{Kobulnicky}, H.~A., {Kennicutt}, Robert~C., J., \& {Pizagno}, J.~L. 1999,
  \hypersetup{urlcolor=magenta}\href{https://dx.doi.org/10.1086/306987}{\apj},
  \hypersetup{urlcolor=blue}\href{https://ui.adsabs.harvard.edu/abs/1999ApJ...514..544K}{514,
  544}

\bibitem[{{Lunnan} {et~al.}(2013){Lunnan}, {Chornock}, {Berger},
  {Milisavljevic}, {Drout}, {Sanders}, {Challis}, {Czekala}, {Foley}, {Fong},
  {Huber}, {Kirshner}, {Leibler}, {Marion}, {McCrum}, {Narayan}, {Rest},
  {Roth}, {Scolnic}, {Smartt}, {Smith}, {Soderberg}, {Stubbs}, {Tonry},
  {Burgett}, {Chambers}, {Kudritzki}, {Magnier}, \& {Price}}]{Lunnan13}
{Lunnan}, R., {Chornock}, R., {Berger}, E., {et~al.} 2013,
  \hypersetup{urlcolor=magenta}\href{https://dx.doi.org/10.1088/0004-637X/771/2/97}{\apj},
  \hypersetup{urlcolor=blue}\href{https://ui.adsabs.harvard.edu/abs/2013ApJ...771...97L}{771,
  97}

\bibitem[{{Lunnan} {et~al.}(2014){Lunnan}, {Chornock}, {Berger}, {Laskar},
  {Fong}, {Rest}, {Sanders}, {Challis}, {Drout}, \& {Foley}}]{Lunnan14}
{Lunnan}, R., {Chornock}, R., {Berger}, E., {et~al.} 2014,
  \hypersetup{urlcolor=magenta}\href{https://dx.doi.org/10.1088/0004-637X/787/2/138}{\apj},
  \hypersetup{urlcolor=blue}\href{https://ui.adsabs.harvard.edu/abs/2014ApJ...787..138L}{787,
  138}

\bibitem[{{Lunnan} {et~al.}(2020){Lunnan}, {Yan}, {Perley}, {Schulze},
  {Taggart}, {Gal-Yam}, {Fremling}, {Soumagnac}, {Ofek}, {Adams}, {Barbarino},
  {Bellm}, {De}, {Fransson}, {Frederick}, {Golkhou}, {Graham}, {Hallakoun},
  {Ho}, {Kasliwal}, {Kaspi}, {Kulkarni}, {Laher}, {Masci}, {Pozo Nu{\~n}ez},
  {Rusholme}, {Quimby}, {Shupe}, {Sollerman}, {Taddia}, {van Roestel}, {Yang},
  \& {Yao}}]{Lunnan20}
{Lunnan}, R., {Yan}, L., {Perley}, D.~A., {et~al.} 2020,
  \hypersetup{urlcolor=magenta}\href{https://dx.doi.org/10.3847/1538-4357/abaeec}{\apj},
  \hypersetup{urlcolor=blue}\href{https://ui.adsabs.harvard.edu/abs/2020ApJ...901...61L}{901,
  61}

\bibitem[{{Lyman} {et~al.}(2016){Lyman}, {Bersier}, {James}, {Mazzali},
  {Eldridge}, {Fraser}, \& {Pian}}]{Lyman16}
{Lyman}, J.~D., {Bersier}, D., {James}, P.~A., {et~al.} 2016,
  \hypersetup{urlcolor=magenta}\href{https://dx.doi.org/10.1093/mnras/stv2983}{\mnras},
  \hypersetup{urlcolor=blue}\href{https://ui.adsabs.harvard.edu/abs/2016MNRAS.457..328L}{457,
  328}

\bibitem[{{Margutti} {et~al.}(2014){Margutti}, {Milisavljevic}, {Soderberg},
  {Chornock}, {Zauderer}, {Murase}, {Guidorzi}, {Sanders}, {Kuin}, {Fransson},
  {Levesque}, {Chandra}, {Berger}, {Bianco}, {Brown}, {Challis},
  {Chatzopoulos}, {Cheung}, {Choi}, {Chomiuk}, {Chugai}, {Contreras}, {Drout},
  {Fesen}, {Foley}, {Fong}, {Friedman}, {Gall}, {Gehrels}, {Hjorth}, {Hsiao},
  {Kirshner}, {Im}, {Leloudas}, {Lunnan}, {Marion}, {Martin}, {Morrell},
  {Neugent}, {Omodei}, {Phillips}, {Rest}, {Silverman}, {Strader},
  {Stritzinger}, {Szalai}, {Utterback}, {Vinko}, {Wheeler}, {Arnett},
  {Campana}, {Chevalier}, {Ginsburg}, {Kamble}, {Roming}, {Pritchard}, \&
  {Stringfellow}}]{Margutti14}
{Margutti}, R., {Milisavljevic}, D., {Soderberg}, A.~M., {et~al.} 2014,
  \hypersetup{urlcolor=magenta}\href{https://dx.doi.org/10.1088/0004-637X/780/1/21}{\apj},
  \hypersetup{urlcolor=blue}\href{https://ui.adsabs.harvard.edu/abs/2014ApJ...780...21M}{780,
  21}

\bibitem[{{Mazzali} {et~al.}(2017){Mazzali}, {Sauer}, {Pian}, {Deng},
  {Prentice}, {Ben Ami}, {Taubenberger}, \& {Nomoto}}]{Mazzali17}
{Mazzali}, P.~A., {Sauer}, D.~N., {Pian}, E., {et~al.} 2017,
  \hypersetup{urlcolor=magenta}\href{https://dx.doi.org/10.1093/mnras/stx992}{\mnras},
  \hypersetup{urlcolor=blue}\href{https://ui.adsabs.harvard.edu/abs/2017MNRAS.469.2498M}{469,
  2498}

\bibitem[{{Milisavljevic} {et~al.}(2015){Milisavljevic}, {Margutti}, {Kamble},
  {Patnaude}, {Raymond}, {Eldridge}, {Fong}, {Bietenholz}, {Challis},
  {Chornock}, {Drout}, {Fransson}, {Fesen}, {Grindlay}, {Kirshner}, {Lunnan},
  {Mackey}, {Miller}, {Parrent}, {Sanders}, {Soderberg}, \&
  {Zauderer}}]{Milisavljevic15}
{Milisavljevic}, D., {Margutti}, R., {Kamble}, A., {et~al.} 2015,
  \hypersetup{urlcolor=magenta}\href{https://dx.doi.org/10.1088/0004-637X/815/2/120}{\apj},
  \hypersetup{urlcolor=blue}\href{https://ui.adsabs.harvard.edu/abs/2015ApJ...815..120M}{815,
  120}

\bibitem[{{Modjaz} {et~al.}(2014){Modjaz}, {Blondin}, {Kirshner}, {Matheson},
  {Berlind}, {Bianco}, {Calkins}, {Challis}, {Garnavich}, {Hicken}, {Jha},
  {Liu}, \& {Marion}}]{Modjaz14}
{Modjaz}, M., {Blondin}, S., {Kirshner}, R.~P., {et~al.} 2014,
  \hypersetup{urlcolor=magenta}\href{https://dx.doi.org/10.1088/0004-6256/147/5/99}{\aj},
  \hypersetup{urlcolor=blue}\href{https://ui.adsabs.harvard.edu/abs/2014AJ....147...99M}{147,
  99}

\bibitem[{{Modjaz} {et~al.}(2020){Modjaz}, {Bianco}, {Siwek}, {Huang},
  {Perley}, {Fierroz}, {Liu}, {Arcavi}, {Gal-Yam}, {Filippenko},
  {Blagorodnova}, {Cenko}, {Kasliwal}, {Kulkarni}, {Schulze}, {Taggart}, \&
  {Zheng}}]{Modjaz20}
{Modjaz}, M., {Bianco}, F.~B., {Siwek}, M., {et~al.} 2020,
  \hypersetup{urlcolor=magenta}\href{https://dx.doi.org/10.3847/1538-4357/ab4185}{\apj},
  \hypersetup{urlcolor=blue}\href{https://ui.adsabs.harvard.edu/abs/2020ApJ...892..153M}{892,
  153}

\bibitem[{{Montero-Dorta} \& {Prada}(2009)}]{montero09}
{Montero-Dorta}, A.~D., \& {Prada}, F. 2009,
  \hypersetup{urlcolor=magenta}\href{https://dx.doi.org/10.1111/j.1365-2966.2009.15197.x}{\mnras},
  \hypersetup{urlcolor=blue}\href{https://ui.adsabs.harvard.edu/abs/2009MNRAS.399.1106M}{399,
  1106}

\bibitem[{{Nadyozhin}(1994)}]{nadyozhin94}
{Nadyozhin}, D.~K. 1994,
  \hypersetup{urlcolor=magenta}\href{https://dx.doi.org/10.1086/192008}{\apjs},
  \hypersetup{urlcolor=blue}\href{https://ui.adsabs.harvard.edu/abs/1994ApJS...92..527N}{92,
  527}

\bibitem[{{Nicholl}(2018)}]{Nicholl18_superbol}
{Nicholl}, M. 2018,
  \hypersetup{urlcolor=magenta}\href{https://dx.doi.org/10.3847/2515-5172/aaf799}{Research
  Notes of the American Astronomical Society},
  \hypersetup{urlcolor=blue}\href{https://ui.adsabs.harvard.edu/abs/2018RNAAS...2d.230N}{2,
  230}

\bibitem[{{Nicholl} {et~al.}(2019){Nicholl}, {Berger}, {Blanchard}, {Gomez}, \&
  {Chornock}}]{Nicholl19_nebular}
{Nicholl}, M., {Berger}, E., {Blanchard}, P.~K., {Gomez}, S., \& {Chornock}, R.
  2019,
  \hypersetup{urlcolor=magenta}\href{https://dx.doi.org/10.3847/1538-4357/aaf470}{\apj},
  \hypersetup{urlcolor=blue}\href{https://ui.adsabs.harvard.edu/abs/2019ApJ...871..102N}{871,
  102}

\bibitem[{{Nicholl} {et~al.}(2017{\natexlab{\hspace{0pt}a}}){Nicholl},
  {Berger}, {Margutti}, {Blanchard}, {Guillochon}, {Leja}, \&
  {Chornock}}]{Nicholl17_egm}
{Nicholl}, M., {Berger}, E., {Margutti}, R., {et~al.}
  2017{\natexlab{\hspace{0pt}a}},
  \hypersetup{urlcolor=magenta}\href{https://dx.doi.org/10.3847/2041-8213/aa82b1}{\apjl},
  \hypersetup{urlcolor=blue}\href{https://ui.adsabs.harvard.edu/abs/2017ApJ...845L...8N}{845,
  L8}

\bibitem[{{Nicholl} {et~al.}(2017{\natexlab{\hspace{0pt}b}}){Nicholl},
  {Guillochon}, \& {Berger}}]{Nicholl17}
{Nicholl}, M., {Guillochon}, J., \& {Berger}, E.
  2017{\natexlab{\hspace{0pt}b}},
  \hypersetup{urlcolor=magenta}\href{https://dx.doi.org/10.3847/1538-4357/aa9334}{\apj},
  \hypersetup{urlcolor=blue}\href{https://ui.adsabs.harvard.edu/abs/2017ApJ...850...55N}{850,
  55}

\bibitem[{{Nicholl} {et~al.}(2016{\natexlab{\hspace{0pt}a}}){Nicholl},
  {Berger}, {Smartt}, {Margutti}, {Kamble}, {Alexander}, {Chen}, {Inserra},
  {Arcavi}, \& {Blanchard}}]{Nicholl16_15bn}
{Nicholl}, M., {Berger}, E., {Smartt}, S.~J., {et~al.}
  2016{\natexlab{\hspace{0pt}a}},
  \hypersetup{urlcolor=magenta}\href{https://dx.doi.org/10.3847/0004-637X/826/1/39}{\apj},
  \hypersetup{urlcolor=blue}\href{https://ui.adsabs.harvard.edu/abs/2016ApJ...826...39N}{826,
  39}

\bibitem[{{Nicholl} {et~al.}(2016{\natexlab{\hspace{0pt}b}}){Nicholl},
  {Berger}, {Margutti}, {Chornock}, {Blanchard}, {Jerkstrand}, {Smartt},
  {Arcavi}, {Challis}, \& {Chambers}}]{Nicholl16_15bnnebular}
{Nicholl}, M., {Berger}, E., {Margutti}, R., {et~al.}
  2016{\natexlab{\hspace{0pt}b}},
  \hypersetup{urlcolor=magenta}\href{https://dx.doi.org/10.3847/2041-8205/828/2/L18}{\apj},
  \hypersetup{urlcolor=blue}\href{https://ui.adsabs.harvard.edu/abs/2016ApJ...828L..18N}{828,
  L18}

\bibitem[{{Nicholl} {et~al.}(2018){Nicholl}, {Blanchard}, {Berger}, {Alexand
  er}, {Metzger}, {Bhirombhakdi}, {Chornock}, {Coppejans}, {Gomez}, \&
  {Margalit}}]{Nicholl18_15bn}
{Nicholl}, M., {Blanchard}, P.~K., {Berger}, E., {et~al.} 2018,
  \hypersetup{urlcolor=magenta}\href{https://dx.doi.org/10.3847/2041-8213/aae70d}{\apj},
  \hypersetup{urlcolor=blue}\href{https://ui.adsabs.harvard.edu/abs/2018ApJ...866L..24N}{866,
  L24}

\bibitem[{{Oliphant}(2007)}]{Oliphant07}
{Oliphant}, T.~E. 2007,
  \hypersetup{urlcolor=magenta}\href{https://dx.doi.org/10.1109/MCSE.2007.58}{CSE},
  \hypersetup{urlcolor=blue}\href{https://ui.adsabs.harvard.edu/abs/2007CSE.....9c..10O}{9,
  10}

\bibitem[{{Pastorello} {et~al.}(2010){Pastorello}, {Smartt}, {Botticella},
  {Maguire}, {Fraser}, {Smith}, {Kotak}, {Magill}, {Valenti}, {Young},
  {Gezari}, {Bresolin}, {Kudritzki}, {Howell}, {Rest}, {Metcalfe}, {Mattila},
  {Kankare}, {Huang}, {Urata}, {Burgett}, {Chambers}, {Dombeck}, {Flewelling},
  {Grav}, {Heasley}, {Hodapp}, {Kaiser}, {Luppino}, {Lupton}, {Magnier},
  {Monet}, {Morgan}, {Onaka}, {Price}, {Rhoads}, {Siegmund}, {Stubbs},
  {Sweeney}, {Tonry}, {Wainscoat}, {Waterson}, {Waters}, \&
  {Wynn-Williams}}]{Pastorello10}
{Pastorello}, A., {Smartt}, S.~J., {Botticella}, M.~T., {et~al.} 2010,
  \hypersetup{urlcolor=magenta}\href{https://dx.doi.org/10.1088/2041-8205/724/1/L16}{\apjl},
  \hypersetup{urlcolor=blue}\href{https://ui.adsabs.harvard.edu/abs/2010ApJ...724L..16P}{724,
  L16}

\bibitem[{{Patat} {et~al.}(2001){Patat}, {Cappellaro}, {Danziger}, {Mazzali},
  {Sollerman}, {Augusteijn}, {Brewer}, {Doublier}, {Gonzalez}, {Hainaut},
  {Lidman}, {Leibundgut}, {Nomoto}, {Nakamura}, {Spyromilio}, {Rizzi},
  {Turatto}, {Walsh}, {Galama}, {van Paradijs}, {Kouveliotou}, {Vreeswijk},
  {Frontera}, {Masetti}, {Palazzi}, \& {Pian}}]{Patat01}
{Patat}, F., {Cappellaro}, E., {Danziger}, J., {et~al.} 2001,
  \hypersetup{urlcolor=magenta}\href{https://dx.doi.org/10.1086/321526}{\apj},
  \hypersetup{urlcolor=blue}\href{https://ui.adsabs.harvard.edu/abs/2001ApJ...555..900P}{555,
  900}

\bibitem[{{Prentice} {et~al.}(2016){Prentice}, {Mazzali}, {Pian}, {Gal-Yam},
  {Kulkarni}, {Rubin}, {Corsi}, {Fremling}, {Sollerman}, {Yaron}, {Arcavi},
  {Zheng}, {Kasliwal}, {Filippenko}, {Cenko}, {Cao}, \& {Nugent}}]{Prentice16}
{Prentice}, S.~J., {Mazzali}, P.~A., {Pian}, E., {et~al.} 2016,
  \hypersetup{urlcolor=magenta}\href{https://dx.doi.org/10.1093/mnras/stw299}{\mnras},
  \hypersetup{urlcolor=blue}\href{https://ui.adsabs.harvard.edu/abs/2016MNRAS.458.2973P}{458,
  2973}

\bibitem[{{Prentice} {et~al.}(2019){Prentice}, {Ashall}, {James}, {Short},
  {Mazzali}, {Bersier}, {Crowther}, {Barbarino}, {Chen}, {Copperwheat},
  {Darnley}, {Denneau}, {Elias-Rosa}, {Fraser}, {Galbany}, {Gal-Yam},
  {Harmanen}, {Howell}, {Hosseinzadeh}, {Inserra}, {Kankare}, {Karamehmetoglu},
  {Lamb}, {Limongi}, {Maguire}, {McCully}, {Olivares E}, {Piascik}, {Pignata},
  {Reichart}, {Rest}, {Reynolds}, {Rodr{\'\i}guez}, {Saario}, {Schulze},
  {Smartt}, {Smith}, {Sollerman}, {Stalder}, {Sullivan}, {Taddia}, {Valenti},
  {Vergani}, {Williams}, \& {Young}}]{Prentice19}
{Prentice}, S.~J., {Ashall}, C., {James}, P.~A., {et~al.} 2019,
  \hypersetup{urlcolor=magenta}\href{https://dx.doi.org/10.1093/mnras/sty3399}{\mnras},
  \hypersetup{urlcolor=blue}\href{https://ui.adsabs.harvard.edu/abs/2019MNRAS.485.1559P}{485,
  1559}

\bibitem[{{Quimby} {et~al.}(2011){Quimby}, {Kulkarni}, {Kasliwal}, {Gal-Yam},
  {Arcavi}, {Sullivan}, {Nugent}, {Thomas}, {Howell}, \& {Nakar}}]{Quimby11}
{Quimby}, R.~M., {Kulkarni}, S.~R., {Kasliwal}, M.~M., {et~al.} 2011,
  \hypersetup{urlcolor=magenta}\href{https://dx.doi.org/10.1038/nature10095}{\nat},
  \hypersetup{urlcolor=blue}\href{https://ui.adsabs.harvard.edu/abs/2011Natur.474..487Q}{474,
  487}

\bibitem[{{Quimby} {et~al.}(2018){Quimby}, {De Cia}, {Gal-Yam}, {Leloudas},
  {Lunnan}, {Perley}, {Vreeswijk}, {Yan}, {Bloom}, \& {Cenko}}]{Quimby18}
{Quimby}, R.~M., {De Cia}, A., {Gal-Yam}, A., {et~al.} 2018,
  \hypersetup{urlcolor=magenta}\href{https://dx.doi.org/10.3847/1538-4357/aaac2f}{\apj},
  \hypersetup{urlcolor=blue}\href{https://ui.adsabs.harvard.edu/abs/2018ApJ...855....2Q}{855,
  2}

\bibitem[{{Roy} {et~al.}(2016){Roy}, {Sollerman}, {Silverman}, {Pastorello},
  {Fransson}, {Drake}, {Taddia}, {Fremling}, {Kankare}, {Kumar}, {Cappellaro},
  {Bose}, {Benetti}, {Filippenko}, {Valenti}, {Nyholm}, {Ergon}, {Sutaria},
  {Kumar}, {Pandey}, {Nicholl}, {Garcia-{\'A}lvarez}, {Tomasella},
  {Karamehmetoglu}, \& {Migotto}}]{Roy16}
{Roy}, R., {Sollerman}, J., {Silverman}, J.~M., {et~al.} 2016,
  \hypersetup{urlcolor=magenta}\href{https://dx.doi.org/10.1051/0004-6361/201527947}{\aap},
  \hypersetup{urlcolor=blue}\href{https://ui.adsabs.harvard.edu/abs/2016A&A...596A..67R}{596,
  A67}

\bibitem[{{Schlafly} \& {Finkbeiner}(2011)}]{Schlafly11}
{Schlafly}, E.~F., \& {Finkbeiner}, D.~P. 2011,
  \hypersetup{urlcolor=magenta}\href{https://dx.doi.org/10.1088/0004-637X/737/2/103}{\apj},
  \hypersetup{urlcolor=blue}\href{https://ui.adsabs.harvard.edu/abs/2011ApJ...737..103S}{737,
  103}

\bibitem[{{Schmidt} {et~al.}(1989){Schmidt}, {Weymann}, \& {Foltz}}]{schmidt89}
{Schmidt}, G.~D., {Weymann}, R.~J., \& {Foltz}, C.~B. 1989,
  \hypersetup{urlcolor=magenta}\href{https://dx.doi.org/10.1086/132495}{\pasp},
  \hypersetup{urlcolor=blue}\href{https://ui.adsabs.harvard.edu/abs/1989PASP..101..713S}{101,
  713}

\bibitem[{{Science Software Branch at STScI}(2012)}]{science12}
{Science Software Branch at STScI}. 2012, {PyRAF: Python alternative for IRAF},
   Astrophysics Source Code Library,
  \hypersetup{urlcolor=magenta}\href{https://ascl.net/1207.011}{ascl}:\hypersetup{urlcolor=blue}\href{https://ui.adsabs.harvard.edu/abs/2012ascl.soft07011S}{1207.011}

\bibitem[{{Smithsonian Astrophysical Observatory}(2000)}]{Smithsonian00}
{Smithsonian Astrophysical Observatory}. 2000, {SAOImage DS9: A utility for
  displaying astronomical images in the X11 window environment},
  \hypersetup{urlcolor=magenta}\href{https://ascl.net/0003.002}{ascl}:\hypersetup{urlcolor=blue}\href{https://ui.adsabs.harvard.edu/abs/2000ascl.soft03002S}{0003.002}

\bibitem[{{Soumagnac} {et~al.}(2020){Soumagnac}, {Ofek}, {Liang}, {Gal-yam},
  {Nugent}, {Yang}, {Cenko}, {Sollerman}, {Perley}, {Andreoni}, {Barbarino},
  {Burdge}, {Bruch}, {De}, {Dugas}, {Fremling}, {Graham}, {Hankins},
  {Strotjohann}, {Moran}, {Neill}, {Schulze}, {Shupe}, {Sip{\H{o}}cz},
  {Taggart}, {Tartaglia}, {Walters}, {Yan}, {Yao}, {Yaron}, {Bellm},
  {Cannella}, {Dekany}, {Duev}, {Feeney}, {Frederick}, {Graham}, {Laher},
  {Masci}, {Kasliwal}, {Kowalski}, {Kupfer}, {Miller}, {Rigault}, \&
  {Rusholme}}]{Soumagnac20}
{Soumagnac}, M.~T., {Ofek}, E.~O., {Liang}, J., {et~al.} 2020,
  \hypersetup{urlcolor=magenta}\href{https://dx.doi.org/10.3847/1538-4357/ab94be}{\apj},
  \hypersetup{urlcolor=blue}\href{https://ui.adsabs.harvard.edu/abs/2020ApJ...899...51S}{899,
  51}

\bibitem[{{Stevenson} {et~al.}(2016){Stevenson}, {Bean}, {Seifahrt}, {Gilbert},
  {Line}, {D{\'e}sert}, \& {Fortney}}]{stevenson16}
{Stevenson}, K.~B., {Bean}, J.~L., {Seifahrt}, A., {et~al.} 2016,
  \hypersetup{urlcolor=magenta}\href{https://dx.doi.org/10.3847/0004-637X/817/2/141}{\apj},
  \hypersetup{urlcolor=blue}\href{https://ui.adsabs.harvard.edu/abs/2016ApJ...817..141S}{817,
  141}

\bibitem[{{Taddia} {et~al.}(2019{\natexlab{\hspace{0pt}a}}){Taddia},
  {Sollerman}, {Fremling}, {Karamehmetoglu}, {Barbarino}, {Lunnan}, {West}, \&
  {Gal-Yam}}]{Taddia19_latetime}
{Taddia}, F., {Sollerman}, J., {Fremling}, C., {et~al.}
  2019{\natexlab{\hspace{0pt}a}},
  \hypersetup{urlcolor=magenta}\href{https://dx.doi.org/10.1051/0004-6361/201833688}{\aap},
  \hypersetup{urlcolor=blue}\href{https://ui.adsabs.harvard.edu/abs/2019A&A...621A..64T}{621,
  A64}

\bibitem[{{Taddia} {et~al.}(2018){Taddia}, {Stritzinger}, {Bersten}, {Baron},
  {Burns}, {Contreras}, {Holmbo}, {Hsiao}, {Morrell}, {Phillips}, {Sollerman},
  \& {Suntzeff}}]{Taddia18}
{Taddia}, F., {Stritzinger}, M.~D., {Bersten}, M., {et~al.} 2018,
  \hypersetup{urlcolor=magenta}\href{https://dx.doi.org/10.1051/0004-6361/201730844}{\aap},
  \hypersetup{urlcolor=blue}\href{https://ui.adsabs.harvard.edu/abs/2018A&A...609A.136T}{609,
  A136}

\bibitem[{{Taddia} {et~al.}(2019{\natexlab{\hspace{0pt}b}}){Taddia},
  {Sollerman}, {Fremling}, {Barbarino}, {Karamehmetoglu}, {Arcavi}, {Cenko},
  {Filippenko}, {Gal-Yam}, \& {Hiramatsu}}]{Taddia19_broadlined}
{Taddia}, F., {Sollerman}, J., {Fremling}, C., {et~al.}
  2019{\natexlab{\hspace{0pt}b}},
  \hypersetup{urlcolor=magenta}\href{https://dx.doi.org/10.1051/0004-6361/201834429}{\aap},
  \hypersetup{urlcolor=blue}\href{https://ui.adsabs.harvard.edu/abs/2019A&A...621A..71T}{621,
  A71}

\bibitem[{{Taubenberger} {et~al.}(2006){Taubenberger}, {Pastorello}, {Mazzali},
  {Valenti}, {Pignata}, {Sauer}, {Arbey}, {B{\"a}rnbantner}, {Benetti}, {Della
  Valle}, {Deng}, {Elias-Rosa}, {Filippenko}, {Foley}, {Goobar}, {Kotak}, {Li},
  {Meikle}, {Mendez}, {Patat}, {Pian}, {Ries}, {Ruiz-Lapuente}, {Salvo},
  {Stanishev}, {Turatto}, \& {Hillebrandt}}]{Taubenberger06}
{Taubenberger}, S., {Pastorello}, A., {Mazzali}, P.~A., {et~al.} 2006,
  \hypersetup{urlcolor=magenta}\href{https://dx.doi.org/10.1111/j.1365-2966.2006.10776.x}{\mnras},
  \hypersetup{urlcolor=blue}\href{https://ui.adsabs.harvard.edu/abs/2006MNRAS.371.1459T}{371,
  1459}

\bibitem[{{van der Walt} {et~al.}(2011){van der Walt}, {Colbert}, \&
  {Varoquaux}}]{Walt11}
{van der Walt}, S., {Colbert}, S.~C., \& {Varoquaux}, G. 2011,
  \hypersetup{urlcolor=magenta}\href{https://dx.doi.org/10.1109/MCSE.2011.37}{CSE},
  \hypersetup{urlcolor=blue}\href{https://ui.adsabs.harvard.edu/abs/2011CSE....13b..22V}{13,
  22}

\bibitem[{{Villar} {et~al.}(2018){Villar}, {Nicholl}, \& {Berger}}]{Villar18}
{Villar}, V.~A., {Nicholl}, M., \& {Berger}, E. 2018,
  \hypersetup{urlcolor=magenta}\href{https://dx.doi.org/10.3847/1538-4357/aaee6a}{\apj},
  \hypersetup{urlcolor=blue}\href{https://ui.adsabs.harvard.edu/abs/2018ApJ...869..166V}{869,
  166}

\bibitem[{{Wang} {et~al.}(2016){Wang}, {Liu}, {Dai}, {Wang}, \& {Wu}}]{Wang16}
{Wang}, S.~Q., {Liu}, L.~D., {Dai}, Z.~G., {Wang}, L.~J., \& {Wu}, X.~F. 2016,
  \hypersetup{urlcolor=magenta}\href{https://dx.doi.org/10.3847/0004-637X/828/2/87}{\apj},
  \hypersetup{urlcolor=blue}\href{https://ui.adsabs.harvard.edu/abs/2016ApJ...828...87W}{828,
  87}

\bibitem[{{Woosley}(2010)}]{Woosley10}
{Woosley}, S.~E. 2010,
  \hypersetup{urlcolor=magenta}\href{https://dx.doi.org/10.1088/2041-8205/719/2/L204}{\apj},
  \hypersetup{urlcolor=blue}\href{https://ui.adsabs.harvard.edu/abs/2010ApJ...719L.204W}{719,
  L204}

\bibitem[{{Woosley} {et~al.}(2007){Woosley}, {Blinnikov}, \&
  {Heger}}]{Woosley07}
{Woosley}, S.~E., {Blinnikov}, S., \& {Heger}, A. 2007,
  \hypersetup{urlcolor=magenta}\href{https://dx.doi.org/10.1038/nature06333}{\nat},
  \hypersetup{urlcolor=blue}\href{https://ui.adsabs.harvard.edu/abs/2007Natur.450..390W}{450,
  390}

\bibitem[{{Yan} {et~al.}(2020{\natexlab{\hspace{0pt}a}}){Yan}, {Perley},
  {Schulze}, {Andreoni}, {Dahiwale}, {Gal-Yam}, {Lunnan}, \&
  {Taggart}}]{Yan20_2019stc}
{Yan}, L., {Perley}, D., {Schulze}, S., {et~al.}
  2020{\natexlab{\hspace{0pt}a}}, Transient Name Server Classification Report,
  \hypersetup{urlcolor=blue}\href{https://ui.adsabs.harvard.edu/abs/2020TNSCR1737....1Y}{2020-1737,
  1}

\bibitem[{{Yan} {et~al.}(2017){Yan}, {Lunnan}, {Perley}, {Gal-Yam}, {Yaron},
  {Roy}, {Quimby}, {Sollerman}, {Fremling}, \& {Leloudas}}]{yan17}
{Yan}, L., {Lunnan}, R., {Perley}, D.~A., {et~al.} 2017,
  \hypersetup{urlcolor=magenta}\href{https://dx.doi.org/10.3847/1538-4357/aa8993}{\apj},
  \hypersetup{urlcolor=blue}\href{https://ui.adsabs.harvard.edu/abs/2017ApJ...848....6Y}{848,
  6}

\bibitem[{{Yan} {et~al.}(2020{\natexlab{\hspace{0pt}b}}){Yan}, {Perley},
  {Schulze}, {Lunnan}, {Sollerman}, {De}, {Chen}, {Fremling}, {Gal-Yam},
  {Taggart}, {Chen}, {Andreoni}, {Bellm}, {Cunningham}, {Dekany}, {Duev},
  {Fransson}, {Laher}, {Hankins}, {Ho}, {Jencson}, {Kaye}, {Kulkarni},
  {Kasliwal}, {Golkhou}, {Graham}, {Masci}, {Miller}, {Neill}, {Ofek},
  {Porter}, {Mr{\'o}z}, {Reiley}, {Riddle}, {Rigault}, {Rusholme}, {Shupe},
  {Soumagnac}, {Smith}, {Tartaglia}, {Yao}, \& {Yaron}}]{Yan20}
{Yan}, L., {Perley}, D.~A., {Schulze}, S., {et~al.}
  2020{\natexlab{\hspace{0pt}b}},
  \hypersetup{urlcolor=magenta}\href{https://dx.doi.org/10.3847/2041-8213/abb8c5}{\apjl},
  \hypersetup{urlcolor=blue}\href{https://ui.adsabs.harvard.edu/abs/2020ApJ...902L...8Y}{902,
  L8}

\bibitem[{{Yaron} \& {Gal-Yam}(2012)}]{Yaron12}
{Yaron}, O., \& {Gal-Yam}, A. 2012,
  \hypersetup{urlcolor=magenta}\href{https://dx.doi.org/10.1086/666656}{\pasp},
  \hypersetup{urlcolor=blue}\href{https://ui.adsabs.harvard.edu/abs/2012PASP..124..668Y}{124,
  668}

\end{thebibliography}

\end{document}